\documentclass[preprint,aps,12pt,notitlepage,nofootinbib,tightenlines]{revtex4}
\usepackage{placeins}
\usepackage{amsmath}
\usepackage{caption}
\usepackage{subcaption}
\usepackage[per-mode=symbol]{siunitx}
\usepackage{booktabs}
\usepackage{multirow}
\usepackage{graphicx}
\usepackage{enumitem}
\usepackage{underscore}
\usepackage{bm}
\usepackage{times}
\usepackage{braket}
\usepackage{color}
\usepackage{epsfig}
\usepackage{slashed}
\usepackage{hyperref}
\usepackage{array}
\usepackage{float}
\newcommand{\beq}{\begin{eqnarray}}
\newcommand{\eeq}{\end{eqnarray}}
\newcommand{\be}{\begin{equation}\begin{aligned}}
\newcommand{\ee}{\end{aligned}\end{equation}}

\definecolor{Red}{rgb}{1.,0.,0.}

\definecolor{Blue}{rgb}{0.,0.,1.}

\definecolor{nicered}{rgb}{0.7,0.1,0.1}
\definecolor{nicegreen}{rgb}{0.1,0.5,0.1}
\def\lsim{ {\ \lower-1.2pt\vbox{\hbox{\rlap{$<$}\lower6pt\vbox{\hbox{$\sim$}}}}\ } }
\def\gsim{ {\ \lower-1.2pt\vbox{\hbox{\rlap{$>$}\lower6pt\vbox{\hbox{$\sim$}}}}\ } }
\hypersetup{colorlinks,citecolor=nicegreen,linkcolor=nicered}
\begin{document}
\title{Probing the pair production  of first-generation vector-like leptons at future $e^+e^-$ colliders}
\author{Yao-Bei Liu$^{1,2}$\footnote{E-mail: liuyaobei@hist.edu.cn}, Stefano Moretti$^{3,4}$\footnote{E-mail: stefano.moretti@cern.ch} }
\affiliation{1. Henan Institute of Science and Technology, Xinxiang 453003, China\\
2. School of Electro-Mechanical Engineering, Zhongyuan Institute of Science and Technology, Xuchang 461000, China\\
3. School of Physics \& Astronomy, University of Southampton, Highfield, Southampton SO17 1BJ, UK \\
4. Department of Physics \& Astronomy, Uppsala University, Box 516, 751 20 Uppsala, Sweden}

\begin{abstract}
This work explores the discovery potential of the first-generation weak isosinglet Vector-Like Leptons (VLLs), denoted by $E^\pm$, via pair production at future electron-positron colliders. Our analysis adopts a comprehensive framework that incorporates beam polarization configurations and leverages detailed detector simulations. We focus on two distinct multilepton signatures: the $2\ell + 2j + \slashed{E}_T$   and  $3\ell + 2j + \slashed{E}_T$ final states ($\ell = e, \mu$). Both signatures arise from the decay  $E^{\pm} \to Z e^{\pm} / W^{\pm} \nu_\ell$ and are distinguished by the decay patterns of the associated gauge bosons. By applying optimized selection criteria to both signal and background events, we establish exclusion sensitivities and discovery prospects across the VLL mass spectrum. Our findings demonstrate that, for integrated luminosities of $\SI{25}{fb^{-1}}$, $\SI{90}{fb^{-1}}$ and $\SI{1000}{fb^{-1}}$ at corresponding center-of-mass
(c.m.) energies of $\SI{1}{TeV}$, $\SI{1.5}{TeV}$ and $\SI{3}{TeV}$, the accessible mass range extends to approximately $\SI{490}{GeV}$, $\SI{740}{GeV}$ and $\SI{1440}{GeV}$, which represents a substantially improvement over the detection limits of existing hadron collider experiments.
\end{abstract}

\maketitle
\newpage
\section{Introduction}
Vector-Like Leptons (VLLs) are among the most promising candidates for physics Beyond the Standard Model (BSM), appearing naturally in many theoretical extensions of the SM. They offer potential solutions to several fundamental problems in particle physics, including the hierarchy problem through new Higgs sector dynamics~\cite{Arkani-Hamed:2012dcq}, viable Dark Matter (DM)   candidates~\cite{Schwaller:2013hqa,Halverson:2014nwa,Bahrami:2016has,Bhattacharya:2018fus}, Electro-Weak (EW) vacuum stabilization~\cite{Xiao:2014kba,Cingiloglu:2024vdh} and explanations for some precision anomalies. The latter encompass the electron and muon $g-2$ discrepancies~\cite{Hiller:2019mou,DeJesus:2020yqx,Frank:2020smf,Dermisek:2021ajd,Brune:2022rlo,Guedes:2022cfy} and observed deviations in the $W^\pm$-boson mass~\cite{He:2022zjz,Kawamura:2022fhm}. The theoretical appeal of VLLs is further strengthened by their natural realization in extra-dimensional scenarios~\cite{Huang:2012kz,Kong:2010qd}, supersymmetric frameworks~\cite{Graham:2009gy,Endo:2011xq,Martin:2012dg,Endo:2012cc,Fischler:2013tva}, Compositeness \cite{DeCurtis:2018iqd} and various other SM extensions~\cite{He:1999vp,Wang:2013jwa,He:2001fz,He:2014ora}.

A defining characteristic of VLLs is indeed their vector-like nature: both left- and right-handed components transform identically under the SM gauge group. This property allows for gauge-invariant mass terms independent of the Higgs mechanism, distinguishing them from chiral SM fermions. Consequently, VLLs face less stringent constraints from EW precision observables~\cite{delAguila:2008pw,Ishiwata:2013gma} and Higgs sector measurements~\cite{Kearney:2012zi,Altmannshofer:2013zba,Falkowski:2013jya}, while generating rich collider phenomenology (see, e.g., Refs.~\cite{Ellis:2014dza,Ishiwata:2015cga,Dermisek:2014cia,Crivellin:2020ebi,Endo:2020tkb,Chakrabarty:2020jro,Guedes:2021oqx,Raju:2022zlv,Li:2023mrw,Dermisek:2014qca,Dermisek:2015oja,Kumar:2015tna,Dermisek:2015hue,Chen:2016lsr,Kawamura:2019rth,Freitas:2020ttd,OsmanAcar:2021plv,Kawamura:2021ygg,Bonilla:2021ize,Baspehlivan:2022qet,Cao:2023smj,Bernreuther:2023uxh,Kawamura:2023zuo,Bigaran:2023ris}).
VLLs that mix with first-, second- or third-generation SM leptons ($\ell = e, \mu, \tau$) are referred to as vector-like electrons, muons or tauons, respectively. This work focuses specifically on the first-generation scenario, so that our VLL is denoted by $E^\pm$, which exhibits unique phenomenological features compared to its heavier counterparts. Experimental searches at the Large Hadron Collider (LHC) have primarily targeted VLL pair production with decays involving third-generation leptons and gauge bosons~\cite{ATLAS:2015qoy,ATLAS:2023sbu,CMS:2019hsm,CMS:2022nty,CMS:2022cpe,CMS:2024bni}. Recent ATLAS results~\cite{ATLAS:2024mrr} establish stringent 95\% Confidence Level (CL) mass limits for vector-like electrons (muons), excluding doublet scenarios up to 1220 GeV (1270 GeV) and singlet scenarios up to 320 GeV (400 GeV). These limits highlight the particular challenge in detecting singlet VLLs at hadron colliders.

In fact, the search for weak-isosinglet VLLs faces significant challenges at hadron colliders because of large QCD backgrounds~\cite{Bhattiprolu:2019vdu}, motivating the exploration of future $e^+e^-$ colliders such as the International
Linear Collider (ILC)~\cite{ILC:2013jhg,ILCInternationalDevelopmentTeam:2022izu} and Compact LInear Collider  (CLIC)~\cite{CLICDetector:2013tfe,Franceschini:2019zsg}, which offer clean experimental conditions for VLL searches~\cite{Yang:2021dtc,Shang:2021mgn,Bhattacharya:2021ltd,Shang:2023rfv,Bhattiprolu:2023yxa,Yue:2024sds,Yue:2024ftz,Shen:2025mxe,Li:2025epjc}. This work examines the pair production of an electron-type weak-isosinglet VLL ($E^\pm$) with minimal mixing to SM electrons at c.m. energies of $\sqrt{s} = \SI{1}{\TeV}$, $\SI{1.5}{\TeV}$ and $\SI{3}{\TeV}$, focusing on the dominant decay channels $E^{\pm} \to W^{\pm}\nu_e$ and $E^{\pm} \to Z^{\pm}e$. By performing comprehensive signal and background simulations, we establish $2\sigma$ exclusion limits and $5\sigma$ discovery reaches, demonstrating that $e^+e^-$ colliders can probe $E^\pm$ masses well beyond current LHC sensitivities, approaching the kinematic threshold at each energy stage and exploring previously inaccessible parameter space.

This paper is organized as follows. Section~\ref{sec:model} introduces the theoretical framework for the singlet VLL model,
in order to specify its interactions with SM particles and to compute the pair-production cross sections at $e^+e^-$ colliders, including the effects of polarized beams. Section~\ref{sec:analysis} presents a detailed collider analysis for the chosen final states, evaluating signals and backgrounds. We conclude with a summary in Section~\ref{sec:conclusion}.
\section{Singlet VLL Model Framework \label{sec:model}}
\subsection{Theoretical setup}
We consider a minimal extension of the SM by introducing a weak isosinglet VLL, denoted \(E^\pm\), as mentioned, which mixes with the first-generation SM leptons. Under the SM gauge group \(SU(3)_C \times SU(2)_L \times U(1)_Y\), the VLL transforms as
\begin{equation}
E_L, E_R^\dagger \sim (\mathbf{1}, \mathbf{1}, -1),
\end{equation}
where both chiral components share the same gauge quantum numbers, permitting a gauge-invariant Dirac mass term.

In the Singlet VLL model for the first generation, the fermion mass terms and
electron mixing with the Standard Model lepton
can be obtained from the Lagrangian
written in 2-component fermion form as
\begin{equation}
-\mathcal{L} = m_{E}E\overline{E} + \epsilon H L_e \overline{E}
           + y_e  H L_e \overline{e} + {\rm c.c.}
 \end{equation}
where $H$ is the SM Higgs complex doublet scalar field,
$L_e = (e, \nu_e)_L$ is the SM first-family lepton doublet
in the gauge eigenstate basis, $y_e$ is the SM electron Yukawa coupling,
and $\epsilon$ is the mixing Yukawa coupling.
The field $E$ ($\overline{E}$) denotes the left-handed (right-handed)
2-component spinor of the vector-like electron.

After EW Symmetry Breaking (EWSB), the mass matrix for the charged leptons in the basis \((e, E)\) is
\begin{equation}
\mathcal{M} = \begin{pmatrix}
y_e v & 0 \\
\epsilon v & M
\end{pmatrix},
\end{equation}
where \(v \approx 174\ \mathrm{GeV}\) is the Higgs Vacuum Expectation Value (VEV). The mixing between the SM electron and the VLL is characterized by a small mixing angle \(\theta_L\), given to Leading Order (LO) by
\begin{equation}
\sin\theta_L \approx \frac{\epsilon v}{M}.
\end{equation}
The physical mass eigenvalues are
\begin{align}
m_e &\approx y_e v \left(1 - \frac{\epsilon^2 v^2}{2 M^2} + \cdots \right), \\
M_E &\approx M \left(1 + \frac{\epsilon^2 v^2}{2 M^2} + \cdots \right).
\end{align}

The mixing angle is severely constrained by current experimental data from EW precision tests, $Z$-pole observables and direct VLL  searches. Recent LHC results~\cite{ATLAS:2024mrr} require $\sin\theta \lesssim 0.03$ for VLL masses in the few hundred GeV range, while low-energy precision measurements impose even tighter constraints at lower masses.
In this work, we adopt a value of $\epsilon = 0.01$. This choice is sufficiently small to enable a perturbative treatment of mixing effects, while being large enough to ensure the prompt decay of
$E^\pm$ within the detector volume~\cite{Kumar:2015tna}. For smaller mixing parameters that induce displaced decays, dedicated searches targeting long-lived charged particles become applicable, as  investigated in Refs.~\cite{Cao:2023smj,Bernreuther:2023uxh}.

\subsection{Pair production at $e^+e^-$ colliders}
The pair production of singlet VLLs proceeds primarily through $s$-channel exchange of EW gauge bosons. The LO interaction Lagrangian for the vector-like electron $E^\pm$, retaining terms up to $\mathcal{O}(\epsilon)$, is given by
\begin{equation}
\mathcal{L}_{\text{int}} = \frac{g s_W^2}{c_W} Z_\mu \left( E^{\dagger} \bar{\sigma}^\mu E - \bar{E}^{\dagger} \bar{\sigma}^\mu \bar{E} \right) - e A_\mu \left( E^{\dagger} \bar{\sigma}^\mu E - \bar{E}^{\dagger} \bar{\sigma}^\mu \bar{E} \right),
\end{equation}
where $e$ denotes the electromagnetic coupling, $g$ is the $SU(2)_L$ gauge coupling, $\theta_{W}$ is the Weinberg angle plus  $s_W \equiv \sin\theta_W$ and  $c_W \equiv \cos\theta_W$ with $e = g s_W$.

For polarized $e^+e^-$ collisions with beam polarizations $P_{e^-}$ and $P_{e^+}$, the partonic cross section takes the form
\begin{equation}
\begin{aligned}
\hat{\sigma}(e^+e^- \rightarrow E^{+}E^{-}) &= \frac{2\pi\alpha^2}{3} (\hat{s} + 2m_{E}^2) \sqrt{1 - 4m_{E}^2/\hat{s}} \\
&\quad \times \left[ |a_L|^2 (1 - P_{e^-})(1 + P_{e^+}) + |a_R|^2 (1 + P_{e^-})(1 - P_{e^+}) \right],
\end{aligned}
\label{eq:cross}
\end{equation}
with the chiral amplitude coefficients
\begin{equation}
a_L = \frac{1}{\hat{s}} + \frac{s_W^2 - 1/2}{c_W^2} \frac{1}{\hat{s} - M_Z^2}, \quad
a_R = \frac{1}{\hat{s}} + \frac{s_W^2}{c_W^2} \frac{1}{\hat{s} - M_Z^2}.
\end{equation}

\begin{figure}[!t]
\begin{center}
\vspace{-0.5cm}
\centerline{\epsfxsize=13cm \epsffile{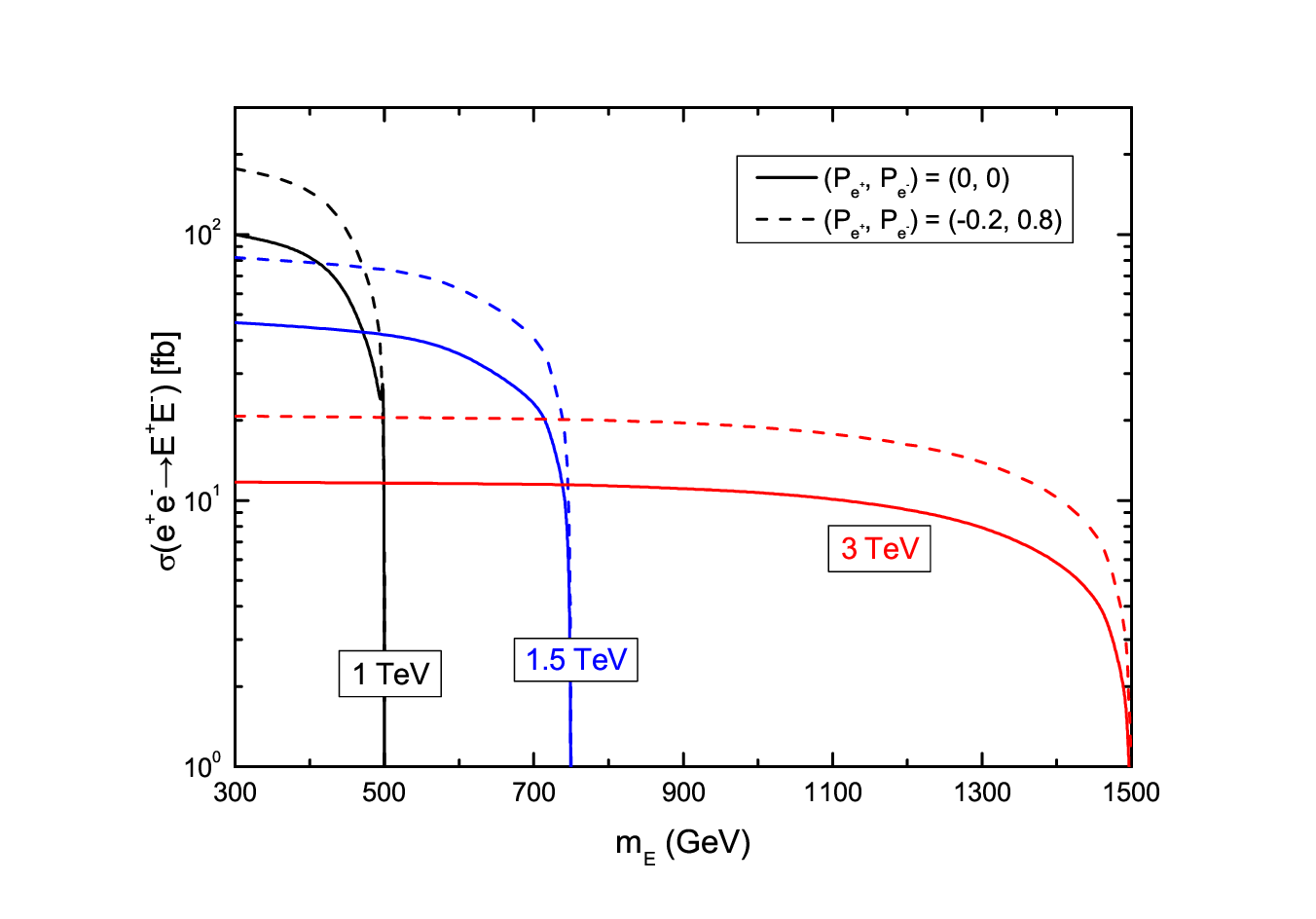}}
\caption{Production cross sections for $e^{+}e^{-}\to E^{+}E^{-}$ as a function of $m_{E}$ at $\sqrt{s} = 1$ TeV, 1.5 TeV and 3 TeV, using the beam polarization configuration $(P_{e^+}, P_{e^-}) = (-0.2, 0.8)$.}
\label{fig:cross}
\end{center}
\end{figure}

Our analysis employs a single optimized beam polarization configuration across all collider scenarios, using $+80\%$ right-handed electrons and $-20\%$ left-handed positrons. This choice follows the technical design reports for future linear colliders~\cite{ILC:2013jhg} and maximizes the production cross section due to the hierarchy $|a_L| < |a_R|$ at high energies. The polarization values adhere to the standard convention where $P = +1$ ($P = -1$) corresponds to a fully right-handed (left-handed) particle.

The production cross sections for $E^+E^-$ pairs, $\sigma(e^+e^-\to E^{+}E^{-})$, are plotted as a function of $m_E$ in Figure~\ref{fig:cross} for three proposed collider configurations: $\sqrt{s} = 1$ TeV (ILC), 1.5 TeV and 3 TeV (CLIC). A distinct threshold behavior is observed, while optimized beam polarization yields a substantial enhancement in the cross section. With such optimization, the cross section reaches approximately 114 fb, 20 fb and 7.8 fb for $m_E$ values of 450 GeV, 740 GeV and 1.45 TeV, respectively, at the corresponding c.m. energies.

\subsection{Decay properties}
The singlet first-generation VLL, $E^\pm$, predominantly decays into SM particles through its small mixing with the SM electron. The main decay channels are the  charged-gauge-current decay $E^{\pm} \to W^{\pm} \nu_e$, the neutral-gauge-current decay $E^{\pm} \to Z^{\pm} e$ and the SM Higgs-mediated decay $E^{\pm} \to h e^{\pm}$.

The corresponding partial decay widths are given by
\begin{align}
\Gamma (E^\pm \rightarrow W^\pm \nu_{e}) &= \frac{\epsilon^2}{32 \pi} m_{E} (1 + 2 r_W) (1 - r_W)^2, \\
\Gamma (E^\pm \rightarrow Z e^\pm) &= \frac{\epsilon^2}{64 \pi} m_{E} (1 + 2 r_Z) (1 - r_Z)^2, \\
\Gamma (E^\pm \rightarrow h e^\pm) &= \frac{\epsilon^2}{64 \pi} m_{E} (1 - r_h)^2,
\end{align}
where $r_X = m_X^2 / m_E^2$ for $X = W^\pm, Z, h$.

In the high-mass regime ($m_E \gg m_W, m_Z, m_h$), the Branching Ratios (BRs) approach the characteristic pattern
\begin{equation}
\text{BR}(E \rightarrow W^\pm \nu_e) : \text{BR}(E \rightarrow Z e^\pm) : \text{BR}(E \rightarrow h e^\pm) = 2 : 1 : 1,
\end{equation}
providing a distinctive signature for VLL identification and discrimination from other BSM scenarios.
\section{Collider Analysis and Signal Sensitivity\label{sec:analysis}}
\begin{figure*}[!t]
\begin{center}
\centerline{\hspace{-1.2cm}\epsfxsize=12cm\epsffile{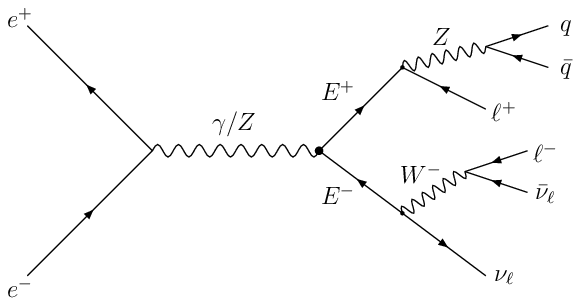}
\hspace{-6.0cm}\epsfxsize=12cm\epsffile{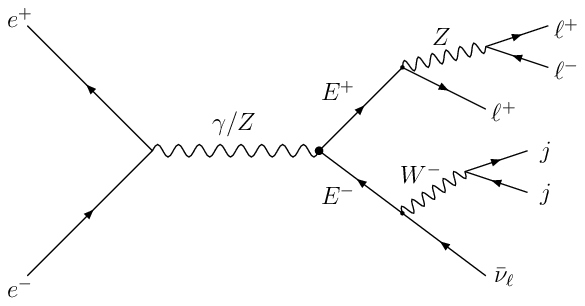}}
\vspace{-12cm}\caption{Characteristic Feynman diagrams illustrating the signal processes for Cases 1 (left) and 2 (right) discussed in the text. }
\label{fey}
\end{center}
\end{figure*}

We assess the discovery potential through comprehensive Monte Carlo (MC)  simulations of both the signal and relevant SM backgrounds. The analysis focuses on the dominant decay channels of the singlet vector-like electron: $E^{\pm} \to W^{\pm}\nu_{e}$ and $E^{\pm} \to Z e^{\pm}$. Considering the subsequent $W$ boson decays $W^{\pm} \to jj$ and $W^{\pm} \to \ell^{\pm}\nu_{\ell}$ as well as  $Z$ boson decays $Z \to \ell^{+}\ell^{-}$ and  $Z \to jj$, the event samples for our search are categorized into the following two topologies:

\begin{itemize}[leftmargin=*,noitemsep]
\item {Case 1}: $e^{+}e^{-} \to E^{+}E^{-} \to Z(\to jj) e^{+}W^{-}(\to \ell^{-}\bar{\nu}_{\ell}) \nu_{e} \to 2\ell + 2j + \slashed{E}_T$;
\item {Case 2}: $e^{+}e^{-} \to E^{+}E^{-} \to Z(\to \ell^{+}\ell^{-}) e^{+}W^{-}(\to jj) \nu_{e} \to 3\ell + 2j + \slashed{E}_T$;
\end{itemize}
\noindent where $\ell = e, \mu$, $j$ denotes light-flavor jets and $\slashed{E}_T$ represents missing transverse energy. The corresponding Feynman diagrams are shown in Figure~\ref{fey}.

Signal and SM background events are generated at LO  using \textsc{MadGraph5\_aMC@NLO}~\cite{mg5}, followed by parton showering and hadronization with \textsc{Pythia}~8.20~\cite{pythia8}. Detector effects are simulated using \textsc{Delphes}~3.4.2~\cite{deFavereau:2013fsa} with the ILD detector card for ILC studies and the CLIC detector card~\cite{Leogrande:2019qbe} for CLIC studies. The event analysis is carried out within the \textsc{MadAnalysis5} framework~\cite{ma5}.

To test the scope of future $e^+e^-$ colliders in testing our VLL scenario, we choose three benchmark configurations of it, with the $E^\pm$ mass spanning values from the  EW to the TeV scale. They are identified as follows, depending on the collider c.m.  energy:
\begin{itemize}
\item $m_E = \SI{350}{GeV}$, \SI{400}{GeV} and \SI{450}{GeV} at $\sqrt{s} = \SI{1}{TeV}$;
\item $m_E = \SI{500}{GeV}$, \SI{600}{GeV} and \SI{700}{GeV} at $\sqrt{s} = \SI{1.5}{TeV}$;
\item $m_E = \SI{800}{GeV}$, \SI{1100}{GeV} and \SI{1400}{GeV} at $\sqrt{s} = \SI{3}{TeV}$.
\end{itemize}

The dominant SM backgrounds include $ZW^{\pm}$, $t\bar{t}Z$ and $t\bar{t}$ production. Table~\ref{cs:sm} summarizes their cross sections including polarization effects at $e^+e^-$ colliders, with ratios $\sigma_{\text{pol}}/\sigma_{\text{unpol}}$ given in parentheses. We find that the adopted beam polarization setup effectively suppresses SM backgrounds, thereby enhancing the sensitivity to our VLL signals.

\begin{table}[ht!]
\centering
\caption{SM background cross sections under the beam polarization configuration $(P_{e^+}, P_{e^-}) = (-0.2, 0.8)$. The values in parentheses denote the cross section ratio $\sigma_{\text{pol}}/\sigma_{\text{unpol}}$. \label{cs:sm}}
\renewcommand{\arraystretch}{1.5}
\begin{tabular}{l c c c}
\hline
\multirow{2}{*}{Process} & \multicolumn{3}{c}{Cross section (fb)} \\
\cline{2-4}
 & $\sqrt{s} = 1$~TeV & $\sqrt{s} = 1.5$~TeV & $\sqrt{s} = 3$~TeV \\
\hline
$e^{+}e^{-} \to ZW^{+}W^{-}$ & 9.95 (0.17) & 10.72 (0.21) & 6.73 (0.21) \\
$e^{+}e^{-} \to t\bar{t}Z$ & 2.70 (0.59) & 1.91 (0.54) & 0.88 (0.53) \\
$e^{+}e^{-} \to t\bar{t}$ & 130 (0.78) & 53 (0.70) & 13.5 (0.70) \\
\hline
\end{tabular}
\end{table}

The object reconstruction adopts the following baseline selection criteria:
\begin{equation}
p_T^{\ell/j} > 10~\text{GeV}, \quad |\eta_{\ell/j}| < 3, \quad \Delta R_{xy} > 0.4,
\end{equation}
where $\Delta R = \sqrt{(\Delta\phi)^2 + (\Delta\eta)^2}$ represents the spatial separation in the $\eta$-$\phi$ plane and the indices $x$ and $y$ refer to different final state particles, such as leptons and jets.

\subsection{Case 1: $2\ell+2j+\slashed{E}_{T}$ analysis}
\begin{figure*}[!t]
\begin{center}
\centerline{\hspace{2.0cm}\epsfxsize=9cm\epsffile{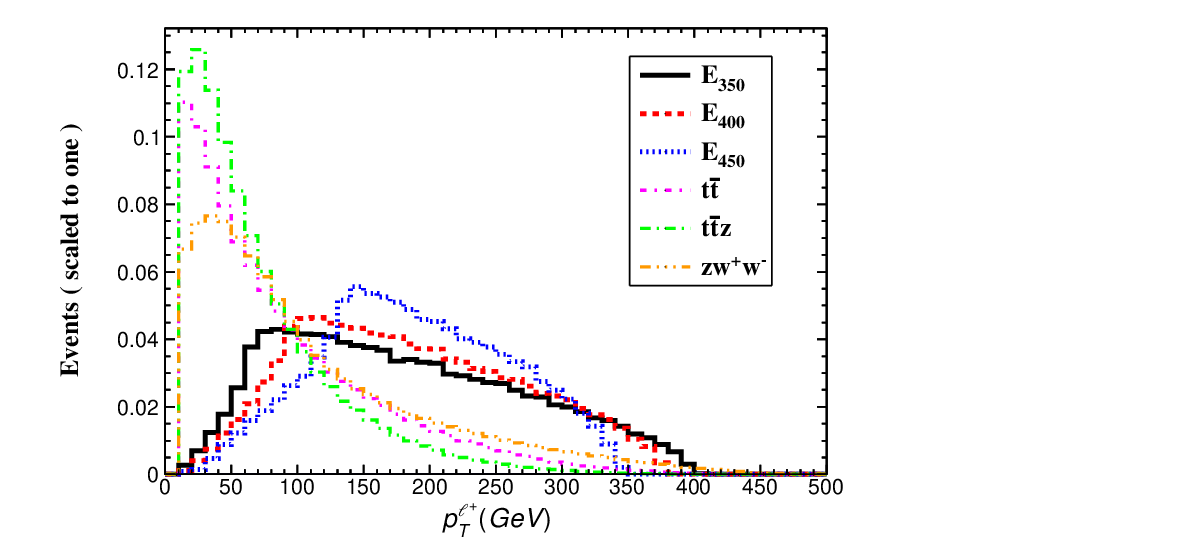}
\hspace{-2.0cm}\epsfxsize=9cm\epsffile{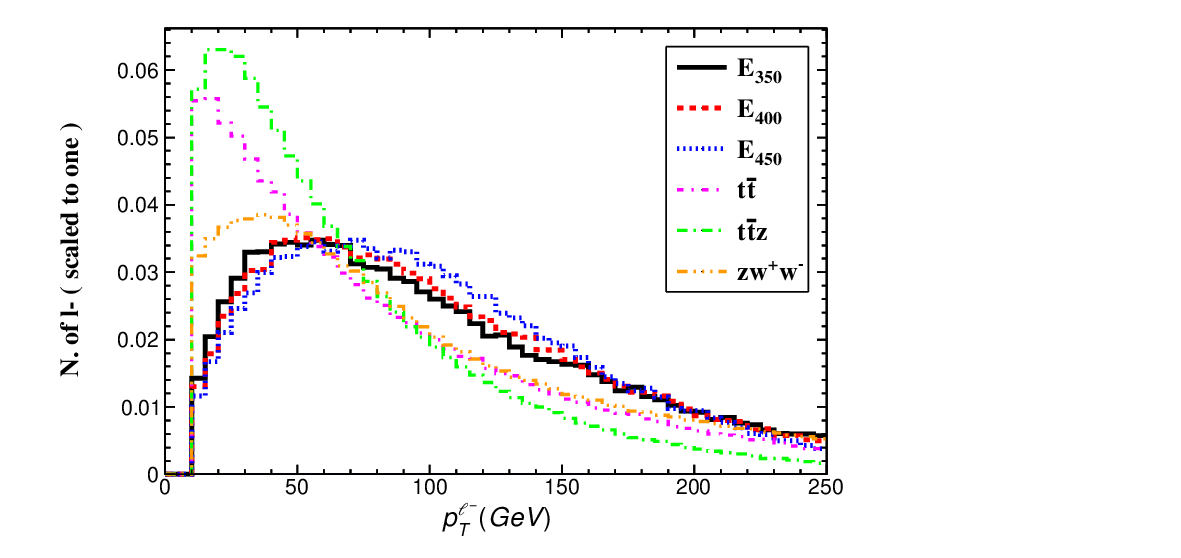}}
\centerline{\hspace{2.0cm}\epsfxsize=9cm\epsffile{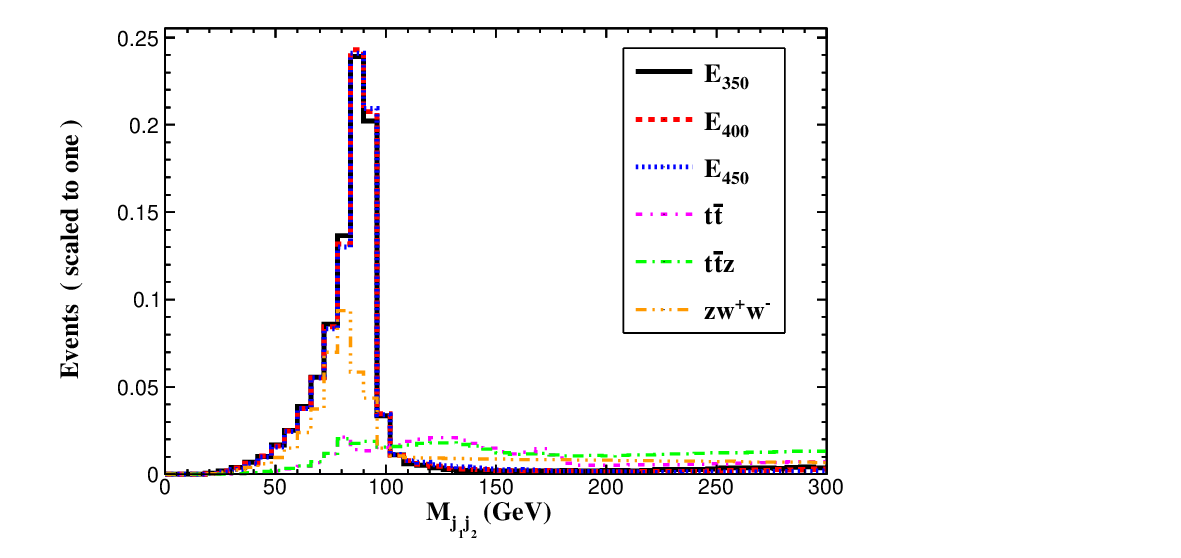}
\hspace{-2.0cm}\epsfxsize=9cm\epsffile{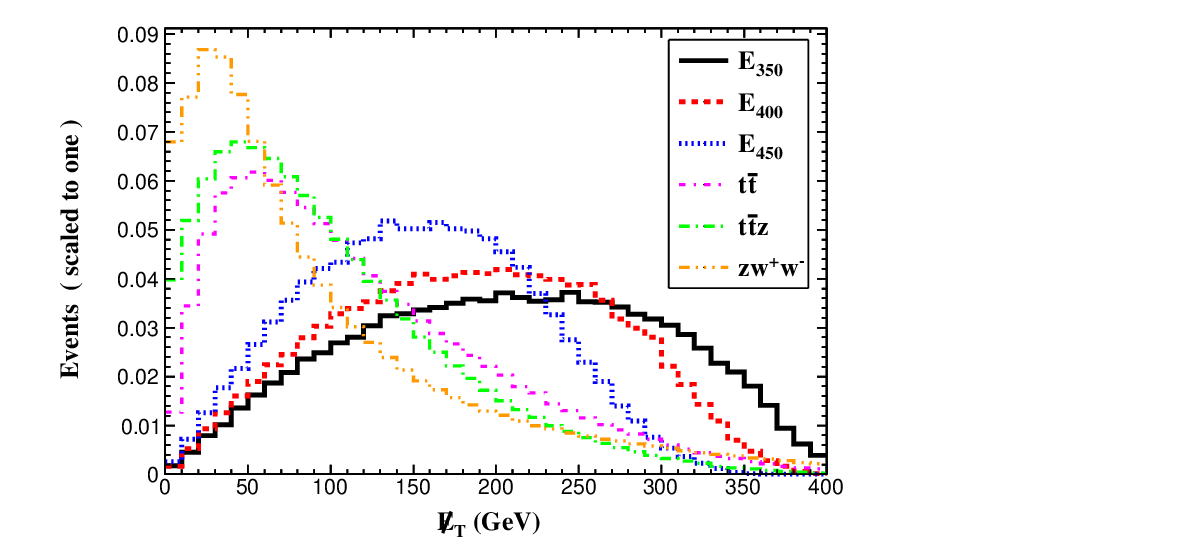}}
\centerline{\hspace{2.0cm}\epsfxsize=9cm\epsffile{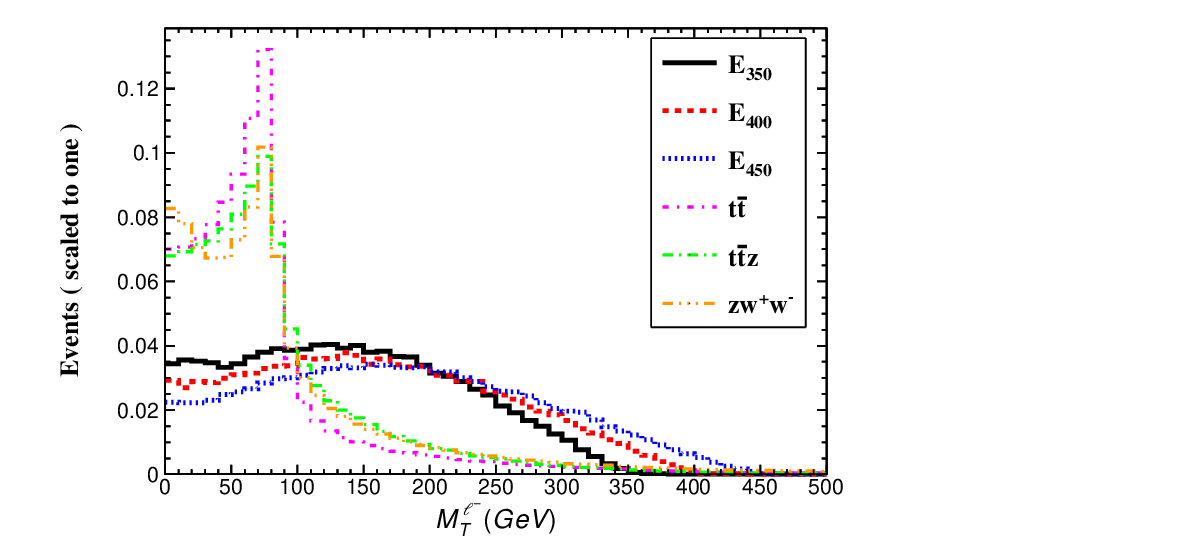}
\hspace{-2.0cm}\epsfxsize=9cm\epsffile{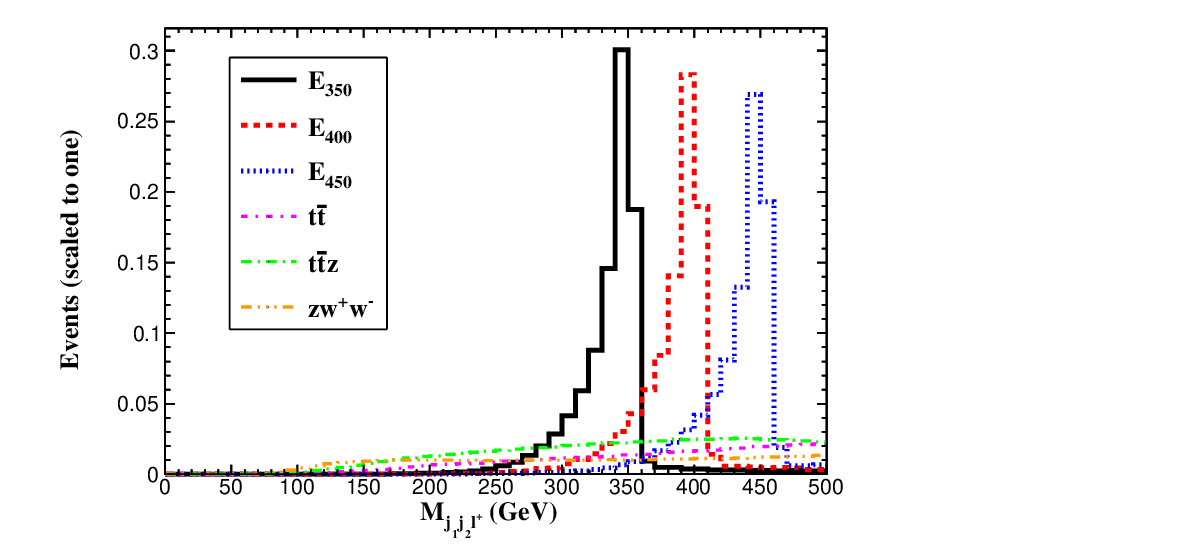}}
\caption{Comparison of normalized distributions between signal processes and relevant SM backgrounds at $\sqrt{s} = 1$ TeV for Case 1. }
\label{fig:distribution1-1}
\end{center}
\end{figure*}
\begin{figure*}[htb]
\begin{center}
\centerline{\hspace{2.0cm}\epsfxsize=9cm\epsffile{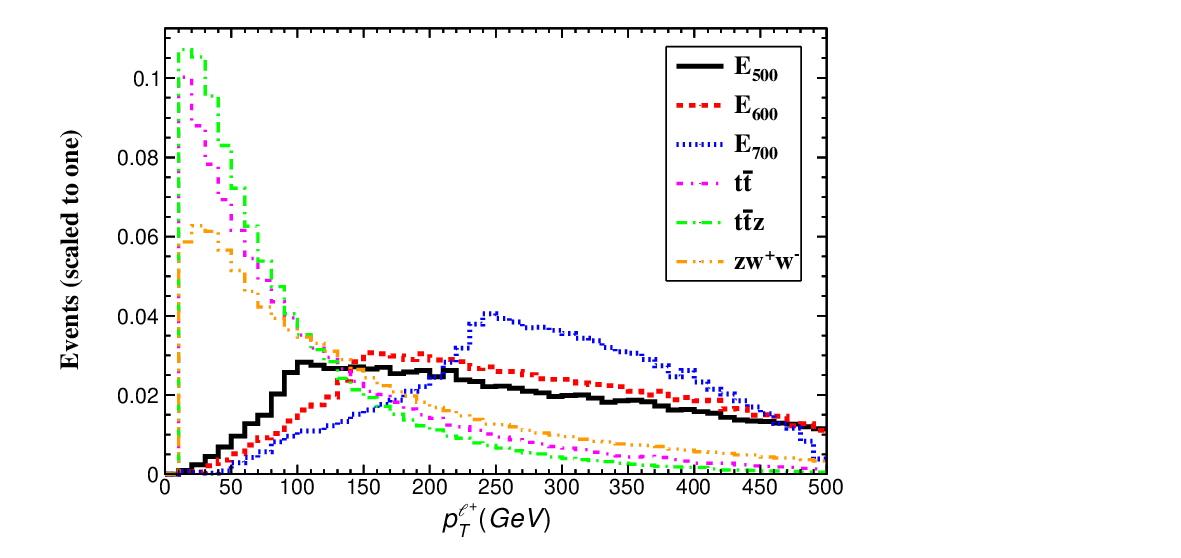}
\hspace{-2.0cm}\epsfxsize=9cm\epsffile{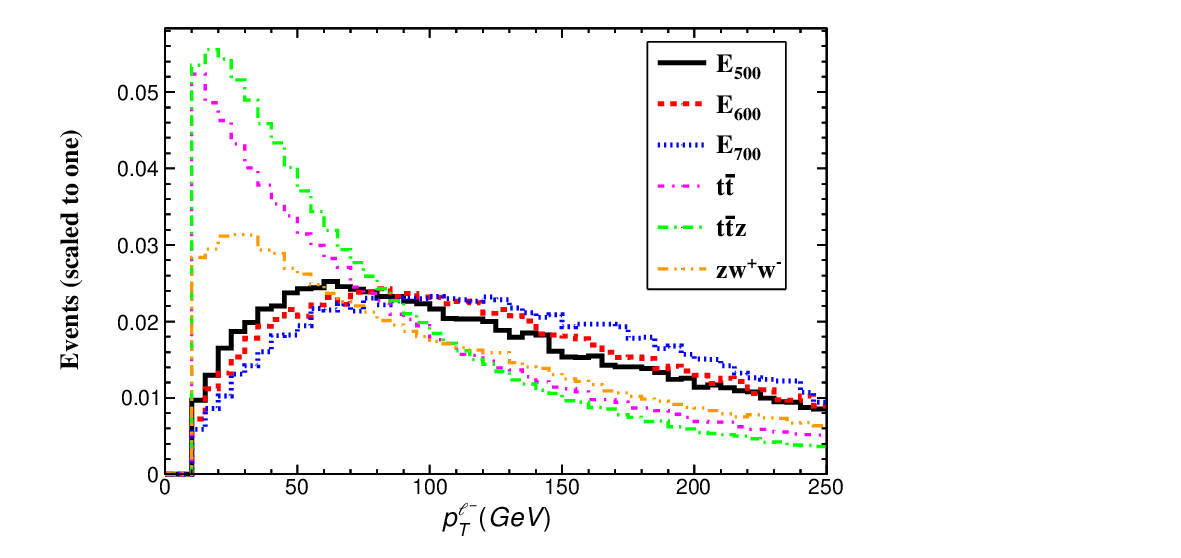}}
\centerline{\hspace{2.0cm}\epsfxsize=9cm\epsffile{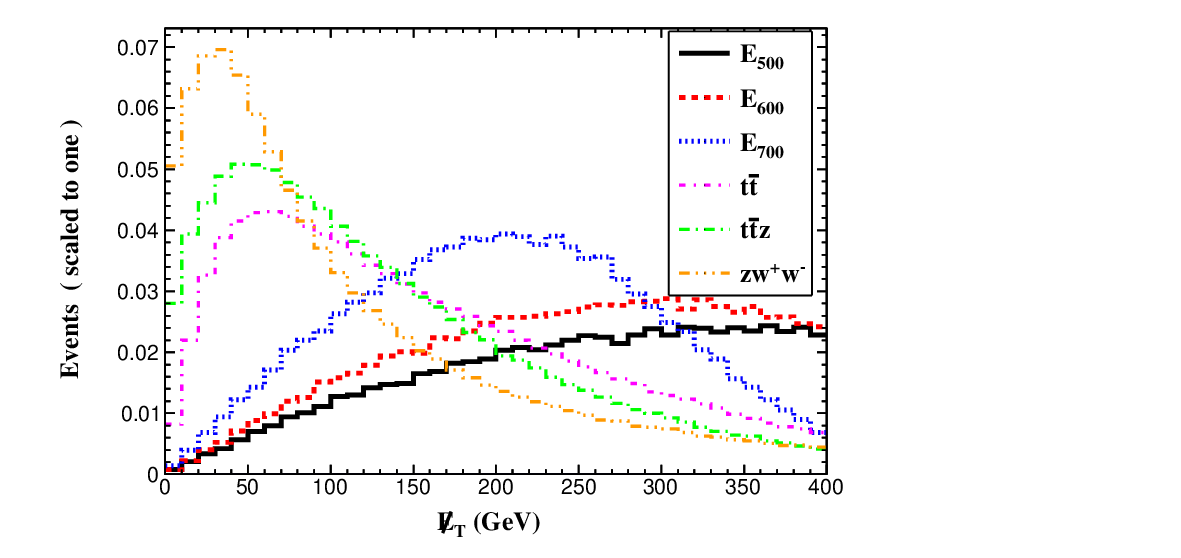}
\hspace{-2.0cm}\epsfxsize=9cm\epsffile{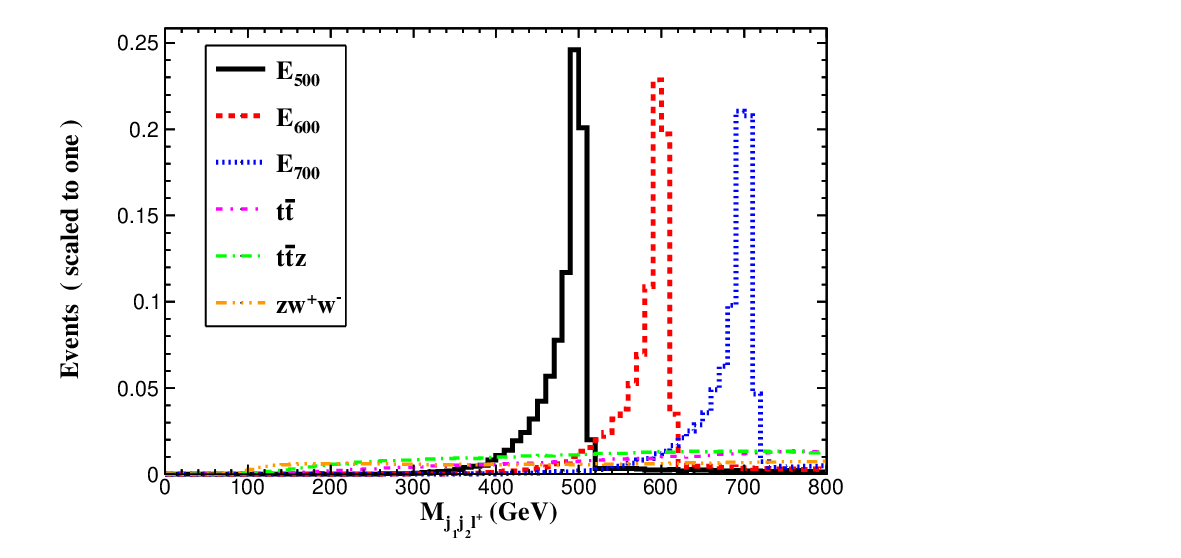}}
\caption{Same as Figure~\ref{fig:distribution1-1}, but for $\sqrt{s} = 1.5$ TeV. }
\label{fig:distribution1-2}
\end{center}
\end{figure*}
\begin{figure*}[htb]
\begin{center}
\centerline{\hspace{2.0cm}\epsfxsize=9cm\epsffile{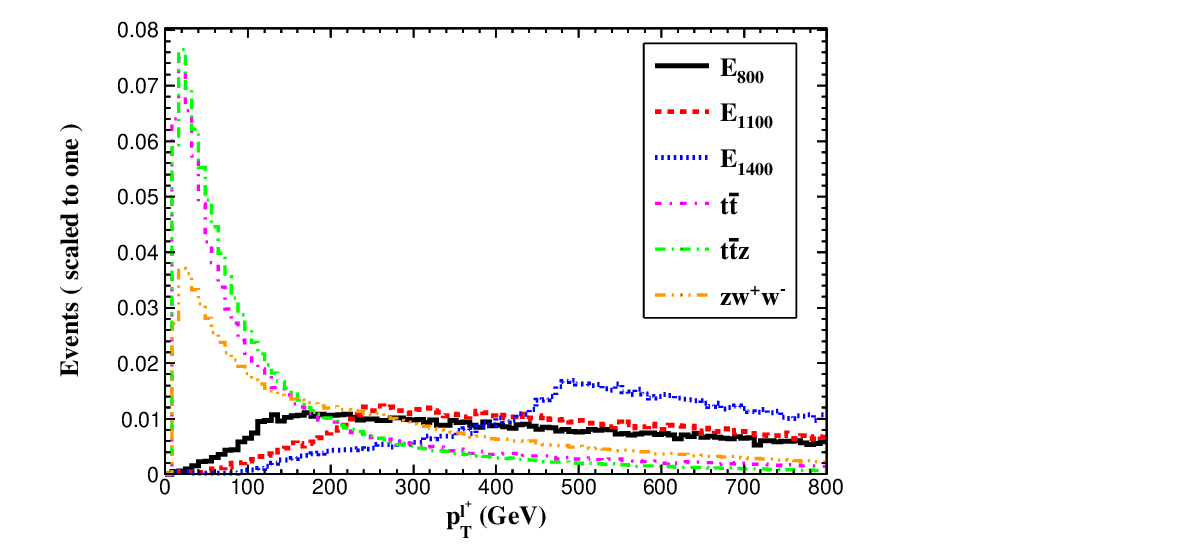}
\hspace{-2.0cm}\epsfxsize=9cm\epsffile{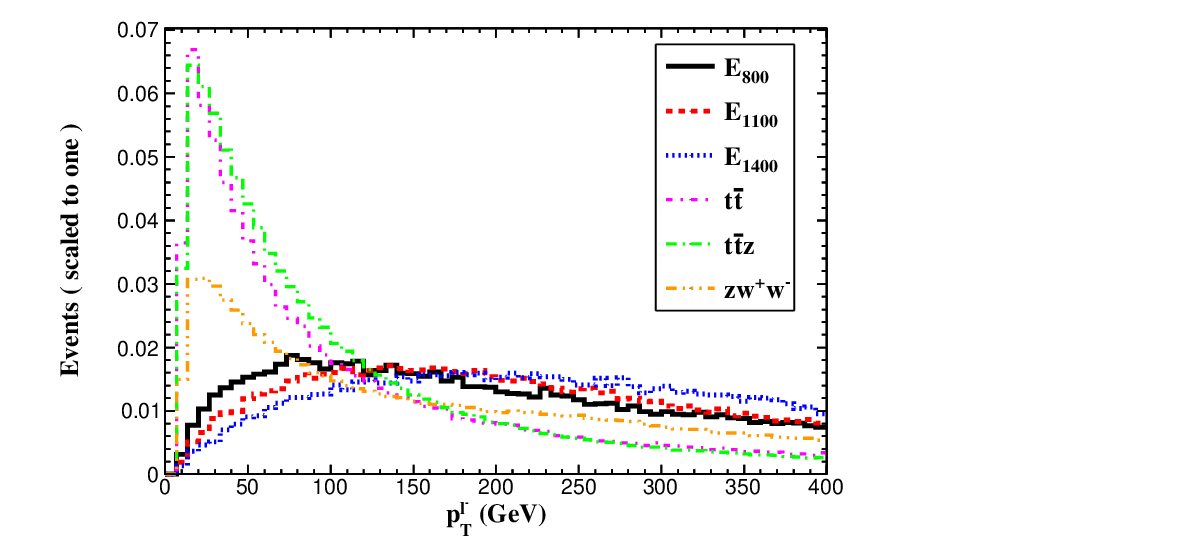}}
\centerline{\hspace{2.0cm}\epsfxsize=9cm\epsffile{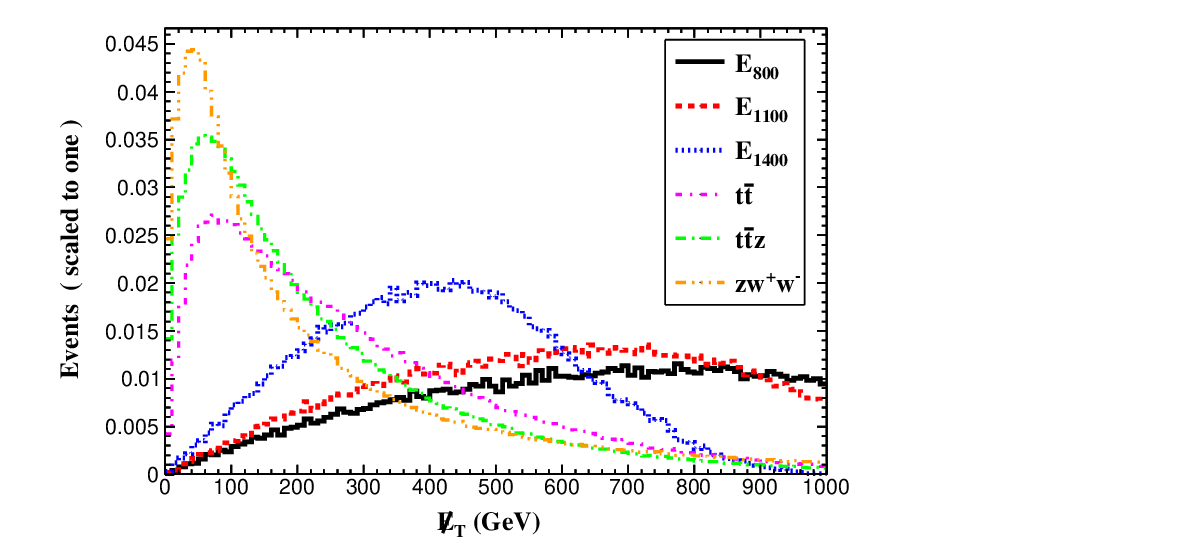}
\hspace{-2.0cm}\epsfxsize=9cm\epsffile{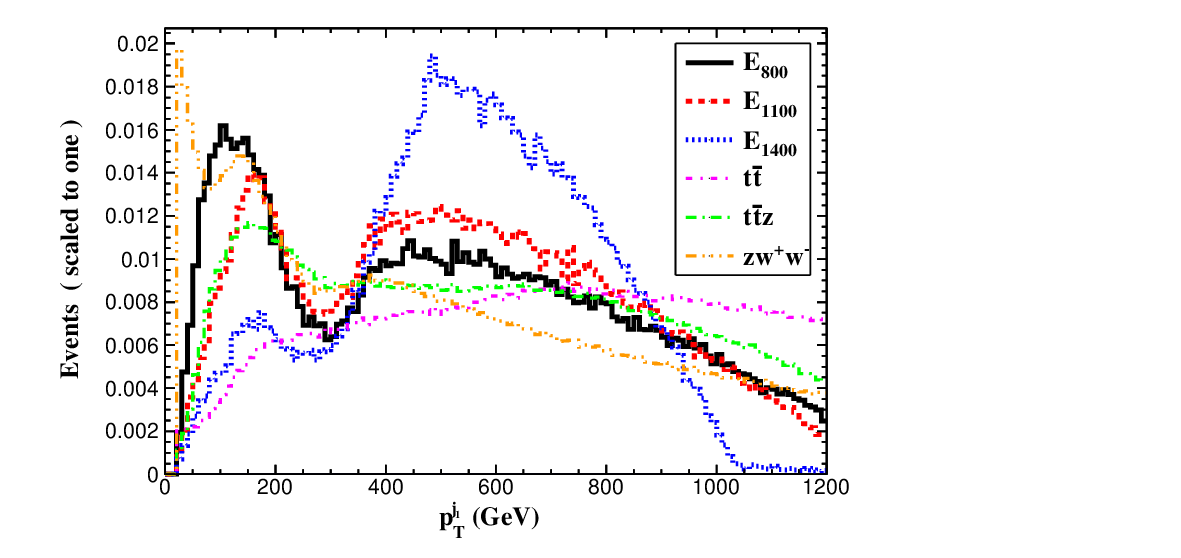}}
\centerline{\hspace{2.0cm}\epsfxsize=9cm\epsffile{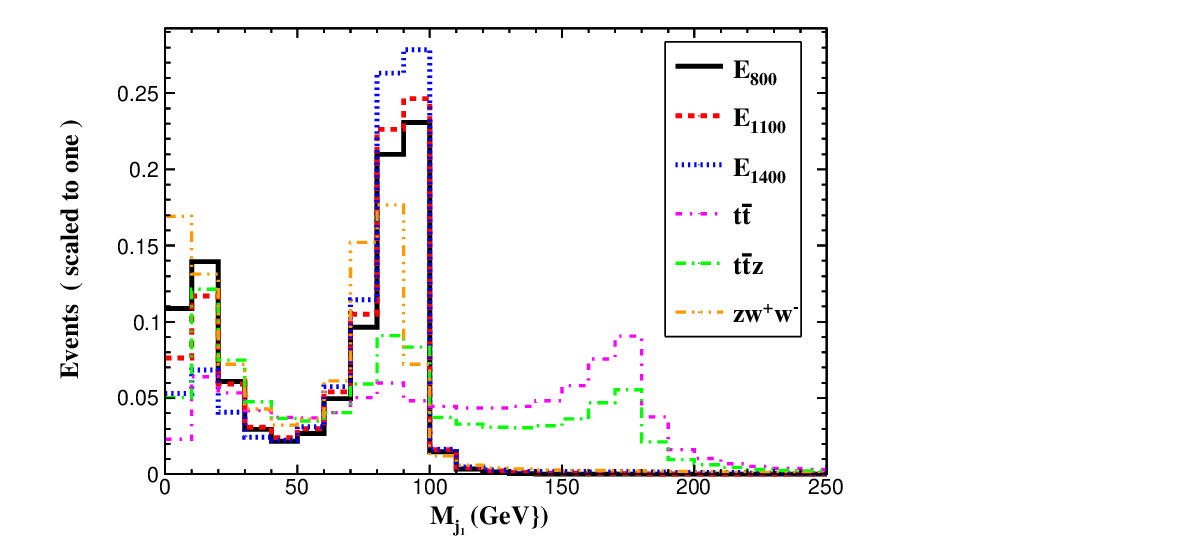}
\hspace{-2.0cm}\epsfxsize=9cm\epsffile{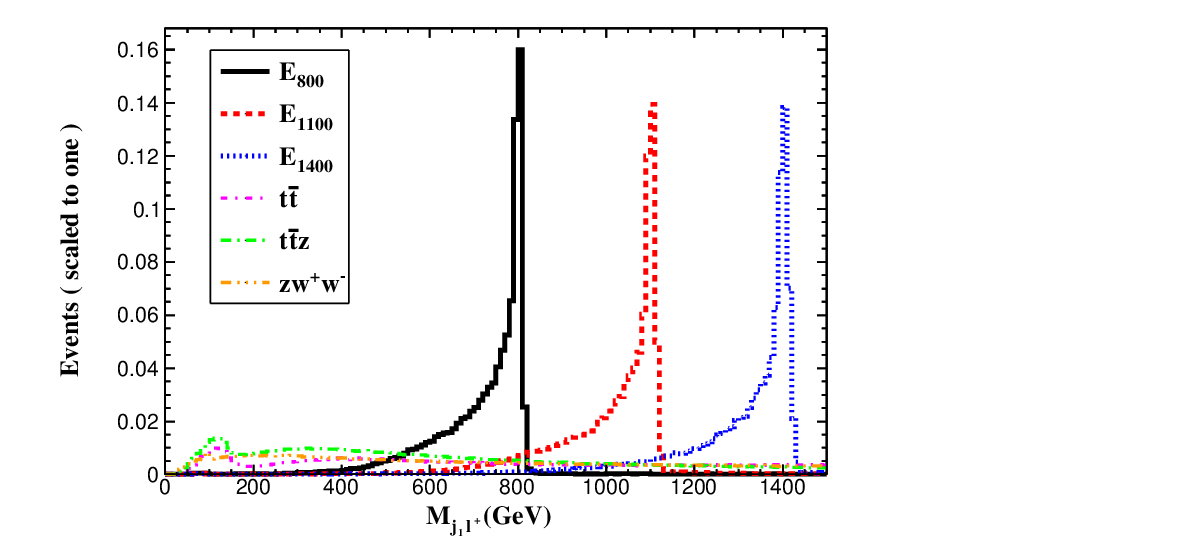}}
\caption{Same as Figure~\ref{fig:distribution1-1}, but for $\sqrt{s} = 3$ TeV. }
\label{fig:distribution1-3}
\end{center}
\end{figure*}
For the $2\ell + 2j + \slashed{E}_{T}$ signal channel (Case~1), the dominant SM backgrounds stem from $e^{+}e^{-} \to ZW^{+}W^{-}$ and $e^{+}e^{-} \to t\bar{t}$ production. Signal events are characterized by two opposite-sign leptons $\ell_1$ and $\ell_2$ that are explicitly non-resonant with the $Z$ boson, along with at least two jets consistent with a $Z$ boson decay. All jets are treated generically, without distinguishing between $b$-jets and light jets. Furthermore, for the decay chain $E^{-} \to W^{-}(\to\ell^-\bar\nu_\ell)\nu_e$ and $E^{+} \to Z(\to jj) e^{+}$ (and charged conjugate), the invariant mass $M_{j_1 j_2 e^{+}}$ can be utilized to reconstruct distinctive mass peaks.

Based on the distinct kinematic features of signal and background events, we show normalized differential distributions in Figures~\ref{fig:distribution1-1}--\ref{fig:distribution1-3} for the aforementioned signal benchmarks.
The distributions are compared with those from the dominant SM backgrounds and include the following observables: the transverse momenta of the two leading leptons $p_T(\ell_1,\ell_2)$, the dijet invariant mass $M_{j_1j_2}$, the missing transverse energy $\slashed{E}_T$, the reconstructed VLL mass $M_{j_1j_2\ell^+}$ and the transverse mass $M_T^{\ell^-}$. While the hadronic $Z$ boson is typically reconstructed from dijet events, in high-energy environments such as the $\sqrt{s} = \SI{3}{TeV}$ scenario, its hadronic decay usually produces a highly boosted single jet rather than a resolved dijet event, prompting the additional inclusion of the transverse momentum $p_T^{j_1}$ and invariant mass $M_{j_1}$ of the leading fat jet. These kinematic observables provide effective discrimination between signal and background processes, particularly through the characteristic VLL mass peaks and distinctive missing transverse energy patterns inherent to the weak-isosinglet VLL signature.

The event selection strategy employs three sequential cut stages.
\begin{itemize}
\item {Cut 1: lepton selection}
\begin{itemize}
\item Selection of exactly two isolated leptons ($e$ or $\mu$) with opposite electric charges.
\item Positive lepton transverse momentum thresholds: $p_T^{\ell_{+}} > \SI{50}{GeV}$ at $\sqrt{s} = \SI{1}{TeV}$, $> \SI{100}{GeV}$ at $\SI{1.5}{TeV}$ and $> \SI{160}{GeV}$ at $\SI{3}{TeV}$.
\item Negative lepton transverse momentum thresholds: $p_T^{\ell_{-}} > \SI{25}{GeV}$ at $\sqrt{s} = \SI{1}{TeV}$, $> \SI{50}{GeV}$ at $\SI{1.5}{TeV}$ and $> \SI{80}{GeV}$ at $\SI{3}{TeV}$.
\item $Z$ boson veto: $|M_{\ell\ell} - m_Z| > \SI{10}{GeV}$.
\end{itemize}

\item {Cut 2: missing transverse energy and transverse mass requirements}
\begin{itemize}
\item Missing transverse energy thresholds: $\slashed{E}_T > \SI{100}{GeV}$ at $\sqrt{s} = \SI{1}{TeV}$, $> \SI{150}{GeV}$ at $\SI{1.5}{TeV}$ and $> \SI{200}{GeV}$ at $\SI{3}{TeV}$.
\item Transverse mass requirement: $M_T(\ell^{-}) > \SI{100}{GeV}$ (applied uniformly across all energy configurations).
\end{itemize}

\item {Cut 3: invariant mass selection}
\begin{itemize}
\item For $\sqrt{s} = \SI{1}{TeV}$ and $\SI{1.5}{TeV}$: requirement of at least two jets with dijet invariant mass $|M_{jj} - m_Z| < \SI{20}{GeV}$, combined with minimum $M_{jj\ell^{+}}$ thresholds of $> \SI{300}{GeV}$ at $\SI{1}{TeV}$ and $> \SI{400}{GeV}$ at $\SI{1.5}{TeV}$.
\item For $\sqrt{s} = \SI{3}{TeV}$: requirement of at least one fat jet with $p_T^{j_1} > \SI{300}{GeV}$ and $\SI{60}{GeV} < M_{j_1} < \SI{110}{GeV}$, along with a minimum $M_{j_1\ell^{+}}$ threshold of $> \SI{600}{GeV}$.
\end{itemize}
\end{itemize}

\begin{table}[htb]
\centering
\footnotesize
\setlength{\tabcolsep}{3pt}
\caption{Cross sections (in fb) for our three signal benchmarks (with different $m_E$) and SM backgrounds in Case 1 at $\sqrt{s} = 1$~TeV. Values in parentheses show the selection efficiency after the corresponding cut is applied.\label{cutflow1-1}}
\begin{tabular}{@{}l cccccc@{}}
\toprule[1.5pt]
\multirow{2}{*}{Cuts} & \multicolumn{3}{c}{Signal ($m_E$ in GeV) } & \multicolumn{3}{c}{Background} \\
\cmidrule(lr){2-4} \cmidrule(lr){5-7}
 & 350 & 400 & 450 & $t\bar{t}$ & $t\bar{t}Z$ & $ZW^{+}W^{-}$ \\
\midrule[0.8pt]
Basic & 2.97 & 2.57 & 1.95 & 40 & 0.97 & 3.63 \\
Cut 1 & 1.86 (61\%) & 1.62 (61\%)  & 1.23 (61\%)  & 1.87 (4.7\%)  & 0.037 (3.8\%)  & 0.20 (5.5\%) \\
Cut 2 & 1.07 (35\%) & 0.98 (37\%)  & 0.76  (38\%)& 0.38 (1.0\%) & 0.008 (0.8\%) & 0.032 (0.9\%) \\
Cut 3 & 0.64 (21\%) & 0.63 (24\%) & 0.52 (26\%)  & 5.4e-3 (0.014\%) & 4.4e-4 (0.045\%) & 0.01 (0.3\%) \\
\bottomrule[1.5pt]
\end{tabular}
\end{table}

\begin{table}[htb]
\centering
\footnotesize
\setlength{\tabcolsep}{3pt}
\caption{Same as Table~\ref{cutflow1-1} but for $\sqrt{s} = 1.5$~TeV. \label{cutflow1-2}}
\begin{tabular}{@{}l cccccc@{}}
\toprule[1.5pt]
\multirow{2}{*}{Cuts} & \multicolumn{3}{c}{Signal ($m_E$ in GeV) } & \multicolumn{3}{c}{Background} \\
\cmidrule(lr){2-4} \cmidrule(lr){5-7}
 & 500 & 600 & 700 & $t\bar{t}$ & $t\bar{t}Z$ & $ZW^{+}W^{-}$ \\
\midrule[0.8pt]
Basic & 1.26 & 1.08 & 0.69 & 12.7 & 0.63 & 3.53 \\
Cut 1 & 0.67 (51\%) & 0.62 (55\%) & 0.41 (56\%) & 0.26 (2.0\%) & 0.01 (1.6\%) & 0.14 (4.0\%) \\
Cut 2 & 0.43 (33\%) & 0.42 (37\%) & 0.27 (37\%) & 0.04 (0.3\%) & 0.002 (0.3\%) & 0.02 (0.6\%) \\
Cut 3 & 0.22 (17\%) & 0.21 (19\%) & 0.15 (21\%) & 5.8e-4 (0.005\%) & 7.8e-5 (0.012\%) & 4.6e-3 (0.13\%) \\
\bottomrule[1.5pt]
\end{tabular}
\end{table}

\begin{table}[htb]
\centering
\footnotesize
\setlength{\tabcolsep}{3pt}
\caption{Same as Table~\ref{cutflow1-1} but for $\sqrt{s} = 3$~TeV.\label{cutflow1-3}}
\begin{tabular}{@{}l cccccc@{}}
\toprule[1.5pt]
\multirow{2}{*}{Cuts} & \multicolumn{3}{c}{Signal ($m_E$ in GeV) } & \multicolumn{3}{c}{Background} \\
\cmidrule(lr){2-4} \cmidrule(lr){5-7}
& 800 & 1100 & 1400 & $t\bar{t}$ & $t\bar{t}Z$ & $ZW^{+}W^{-}$ \\
\midrule[0.8pt]
Basic & 0.32 & 0.28 & 0.17 & 2.82 & 0.28 & 2.15 \\
Cut 1 & 0.14 (41\%) & 0.14 (47\%) & 0.086 (49\%) & 0.037 (1.3\%) & 0.003 (1.1\%) & 0.074 (3.4\%) \\
Cut 2 & 0.10 (30\%) & 0.10 (35\%) & 0.058 (33\%) & 0.0083 (0.29\%) & 0.00078 (0.28\%) & 0.011 (0.51\%) \\
Cut 3 & 0.058 (17\%) & 0.068 (23\%) & 0.046 (26\%) & 8.6e-4 (0.03\%) & 2.2e-4 (0.079\%) & 5.1e-3 (0.24\%) \\
\bottomrule[1.5pt]
\end{tabular}
\end{table}
The cutflow results for our signal benchmarks with the three chosen $E$ masses
and dominant SM backgrounds are summarized in Tables~\ref{cutflow1-1}-\ref{cutflow1-3} for c.m. energies of \SI{1}{TeV}, \SI{1.5}{TeV} and \SI{3}{TeV}. The cross sections after each sequential selection are computed as $\sigma_{\text{cut}} = \sigma_0 \times \epsilon_{\text{cut}}$, where $\sigma_0$ denotes the initial cross section and $\epsilon_{\text{cut}}$ represents the cumulative efficiency up to that stage. The efficiency for each individual cut is determined from the ratio of successive cross section values. Upon completion of the full selection procedure, the total SM background cross sections are suppressed to $\SI{3.3e-3}{fb}$ at $\sqrt{s} = \SI{1}{TeV}$, $\SI{1.5e-3}{fb}$ at $\sqrt{s} = \SI{1.5}{TeV}$ and $\SI{0.9e-3}{fb}$ at $\sqrt{s} = \SI{3}{TeV}$, confirming efficient background rejection across all considered collision energies.

\subsection{Case 2: $3\ell + 2j + \slashed{E}_T$ analysis}
The dominant SM backgrounds for the $3\ell + 2j + \slashed{E}_{T}$ final state (Case~2) are $e^{+}e^{-} \to ZW^{+}W^{-}$ and $e^{+}e^{-} \to t\bar{t}Z$. In signal events, the highest-$p_T$ isolated lepton ($\ell_1$) originates from the decay $E^\pm \to  Ze^\pm$, while an Opposite-Sign Same-Flavor (OSSF) lepton pair ($\ell_2$, $\ell_3$) comes from the $Z$-boson decay whereas the two jets arise predominantly from hadronic $W^\pm$-boson decays. Reconstructible mass peaks emerge in $M_{\ell_1 \ell_2 \ell_3}$ and $M_T^{j_1 j_2}$ for the processes $E^{\mp} \to W^{\mp}\nu_e$ and $E^{\pm} \to Z e^{\pm}$. At $\sqrt{s} = \SI{3}{TeV}$, the $W$ boson from heavy $E^\pm$ decay is highly boosted, causing its hadronic decay products to merge into a single fat jet; accordingly, the transverse momentum and invariant mass distributions of the leading fat jet $j_1$ are included in the analysis.

\begin{figure*}[!t]
\begin{center}
\centerline{\hspace{2.0cm}\epsfxsize=9cm\epsffile{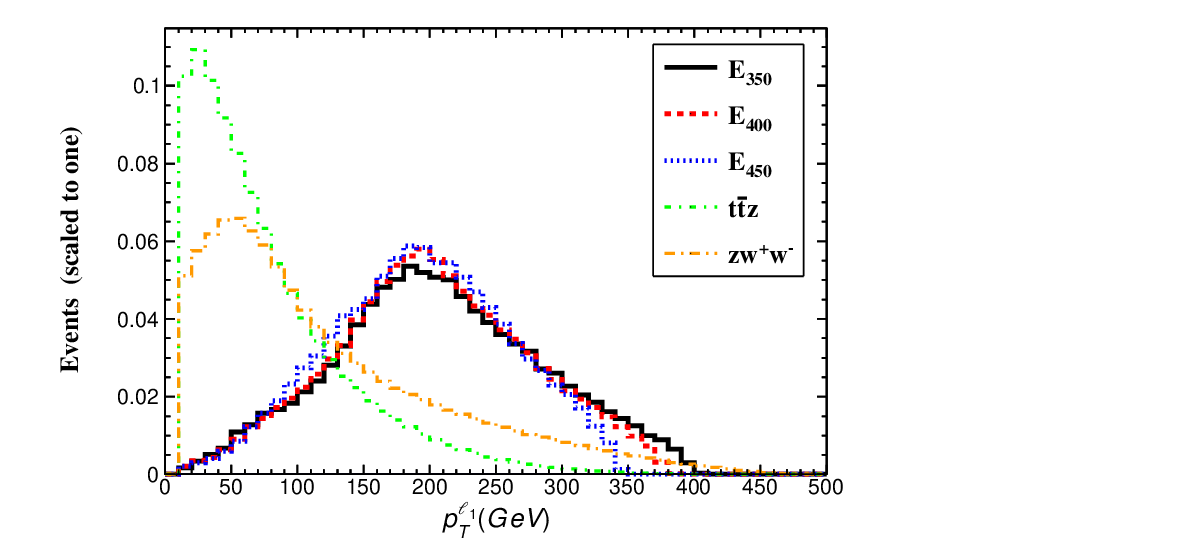}
\hspace{-2.0cm}\epsfxsize=9cm\epsffile{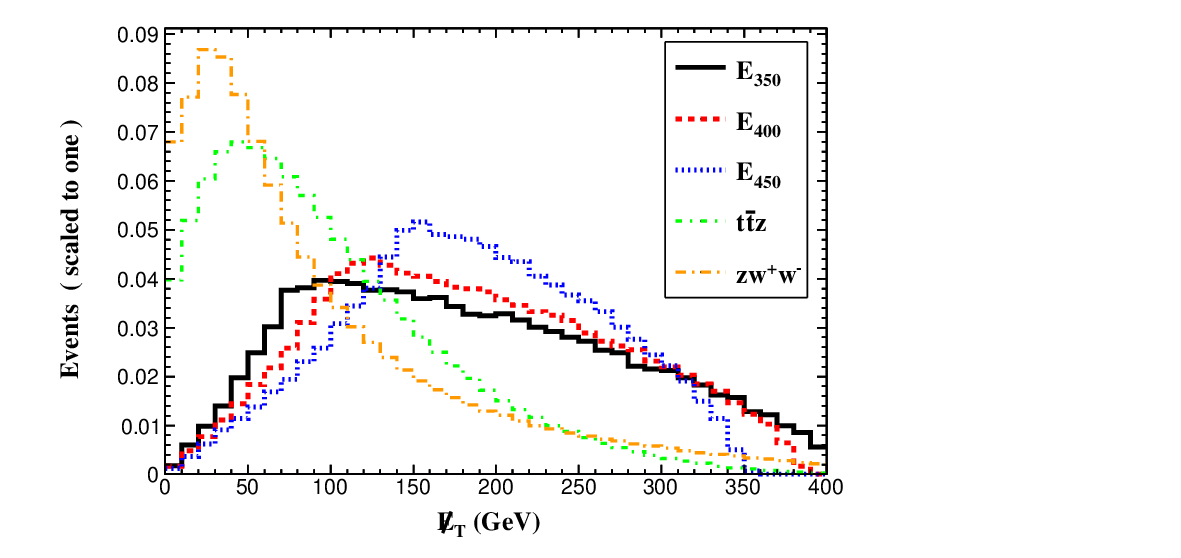}}
\centerline{\hspace{2.0cm}\epsfxsize=9cm\epsffile{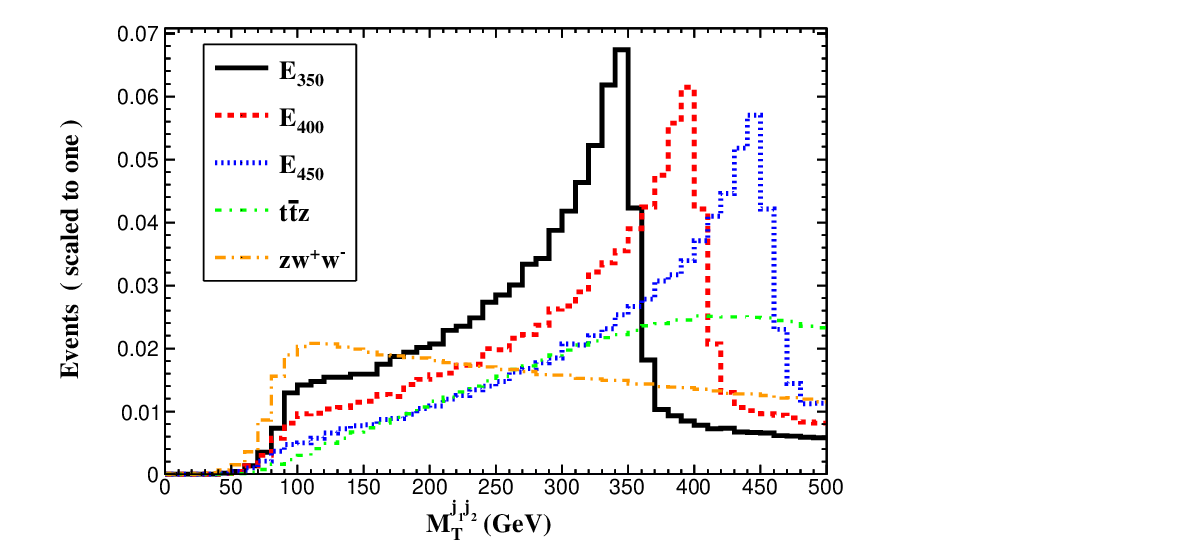}
\hspace{-2.0cm}\epsfxsize=9cm\epsffile{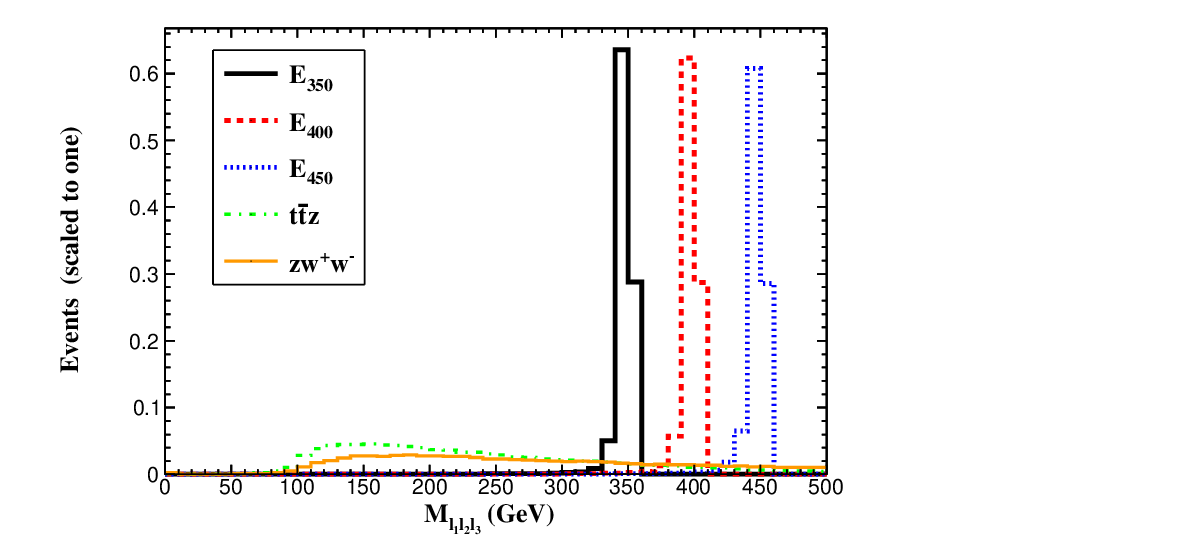}}
\caption{Normalized distributions for the signals and relevant SM backgrounds at $\sqrt{s} = 1$ TeV for Case 2. }
\label{fig:distribution2-1}
\end{center}
\end{figure*}
\begin{figure*}[htb]
\begin{center}
\centerline{\hspace{2.0cm}\epsfxsize=9cm\epsffile{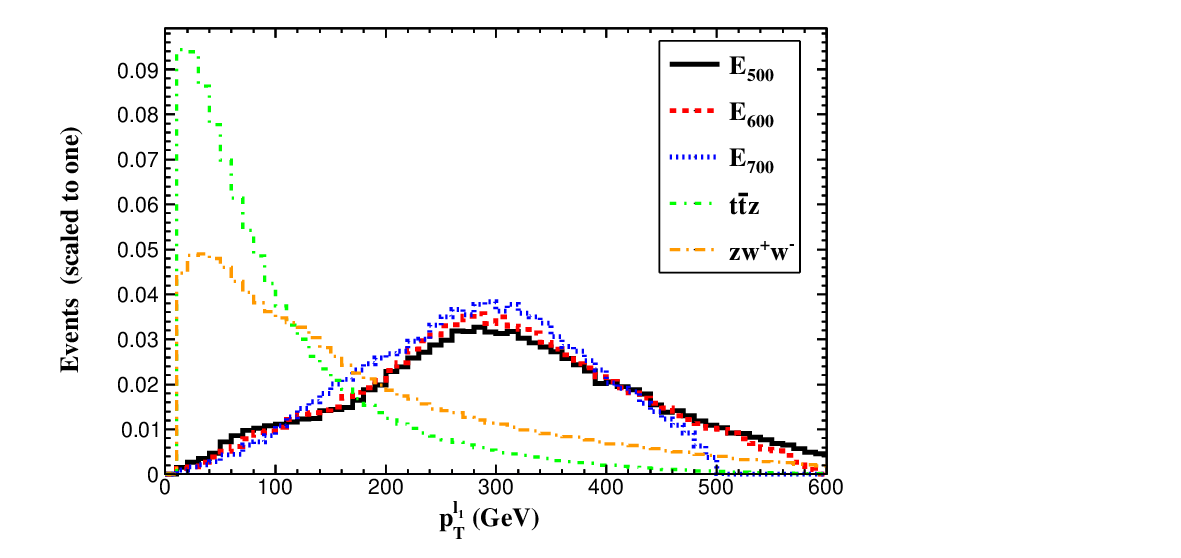}
\hspace{-2.0cm}\epsfxsize=9cm\epsffile{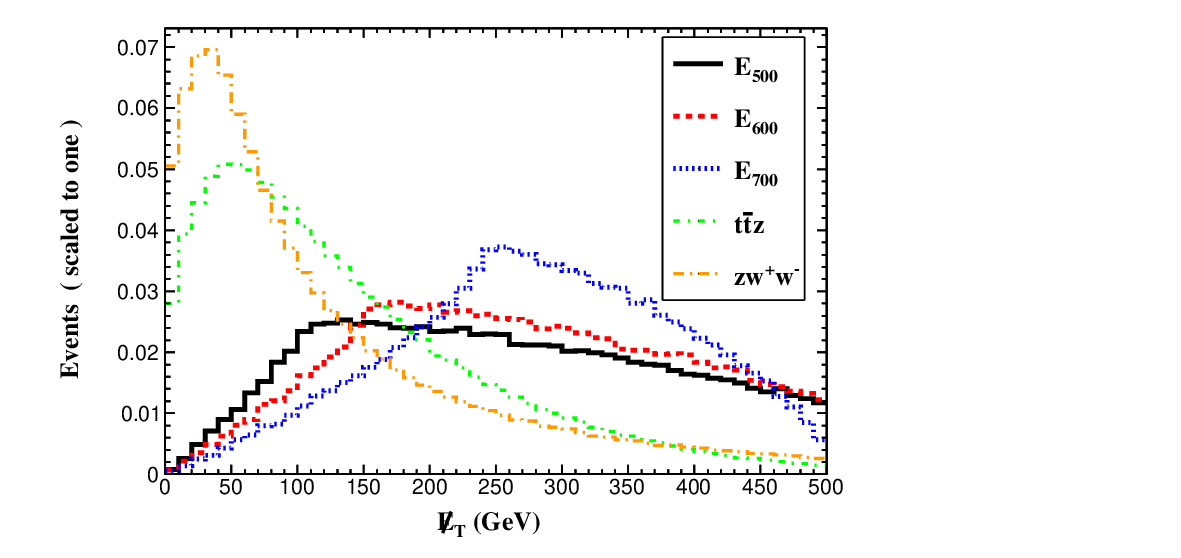}}
\centerline{\hspace{2.0cm}\epsfxsize=9cm\epsffile{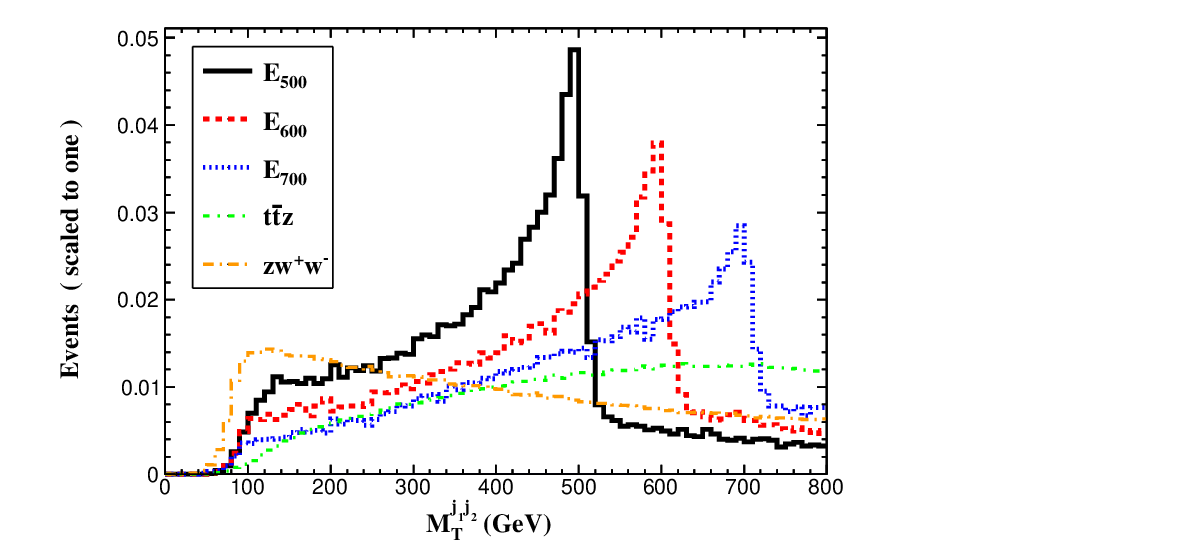}
\hspace{-2.0cm}\epsfxsize=9cm\epsffile{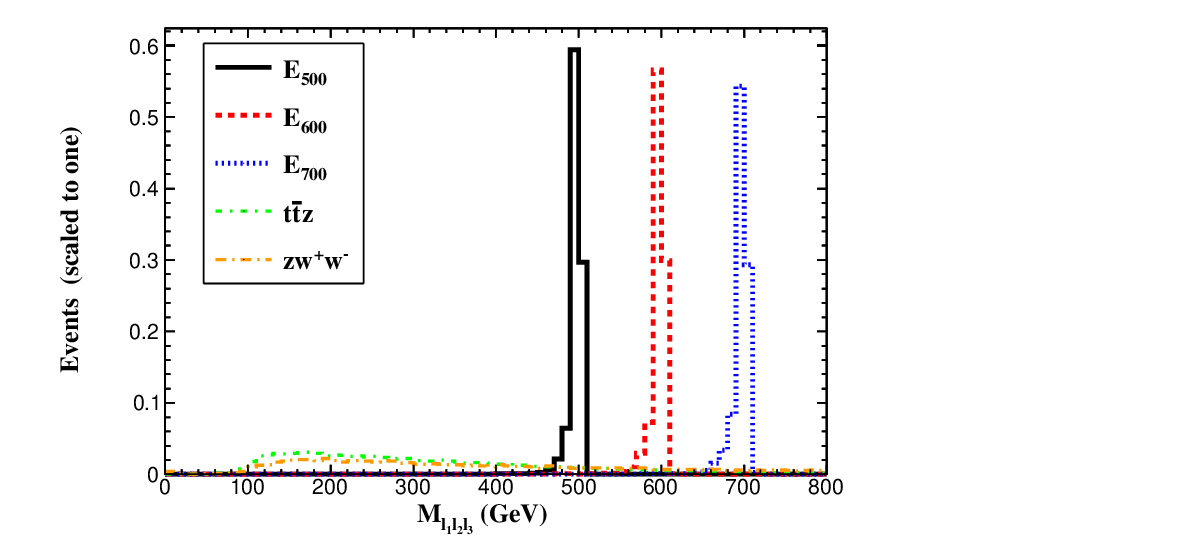}}
\caption{Same as Figure~\ref{fig:distribution2-1}, but for $\sqrt{s} = 1.5$ TeV.  }
\label{fig:distribution2-2}
\end{center}
\end{figure*}
\begin{figure*}[htb]
\begin{center}
\centerline{\hspace{2.0cm}\epsfxsize=9cm\epsffile{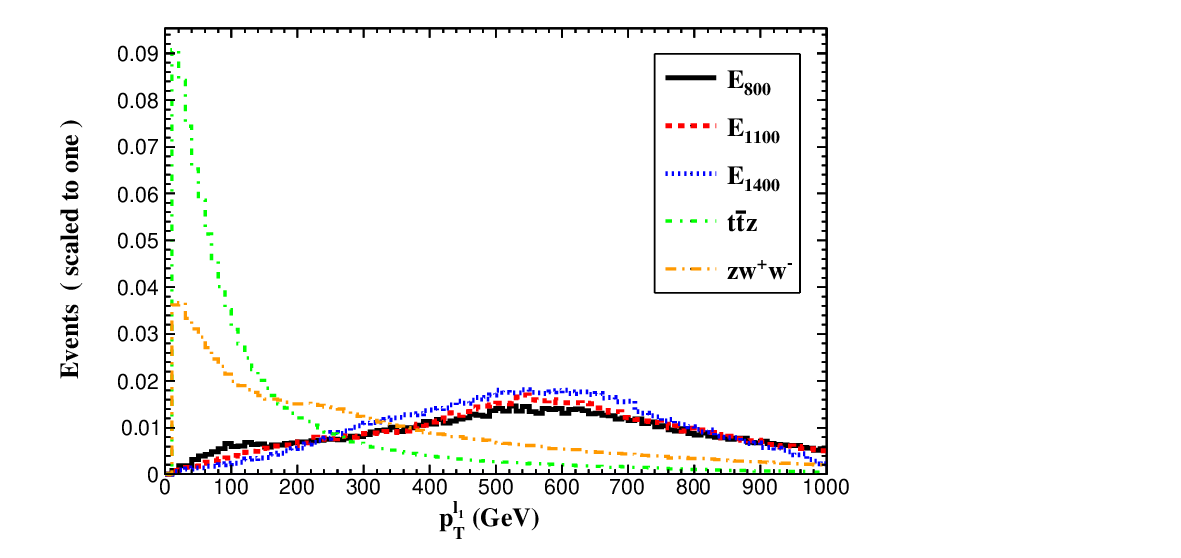}
\hspace{-2.0cm}\epsfxsize=9cm\epsffile{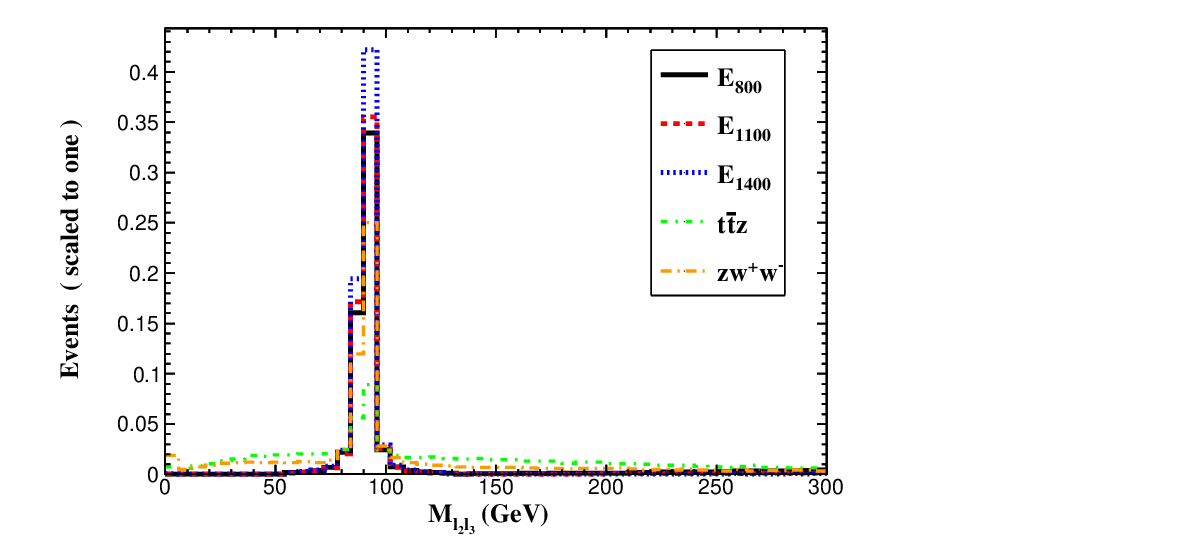}}
\centerline{\hspace{2.0cm}\epsfxsize=9cm\epsffile{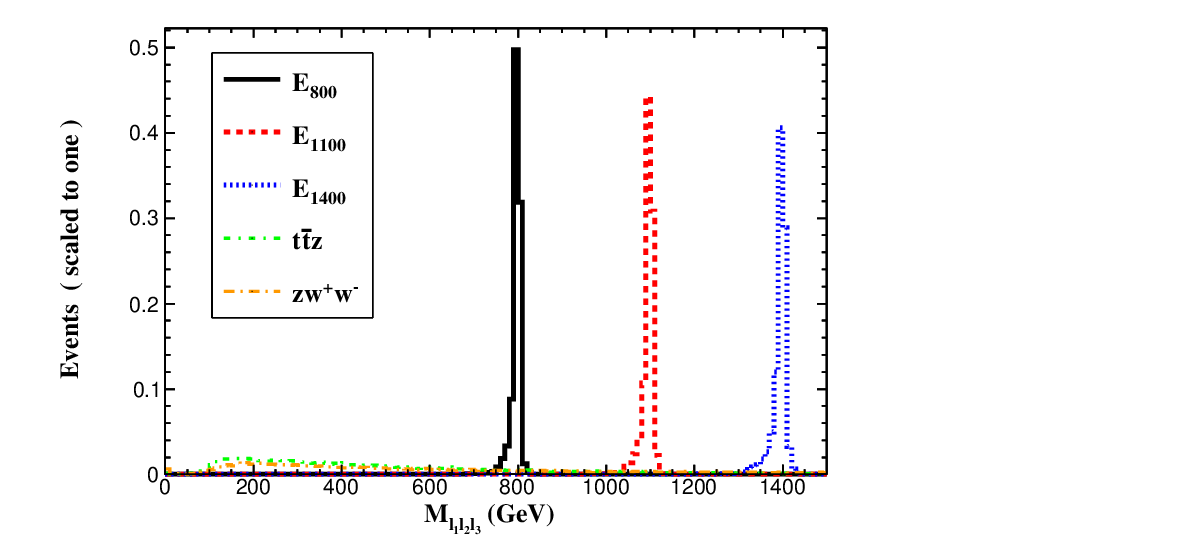}
\hspace{-2.0cm}\epsfxsize=9cm\epsffile{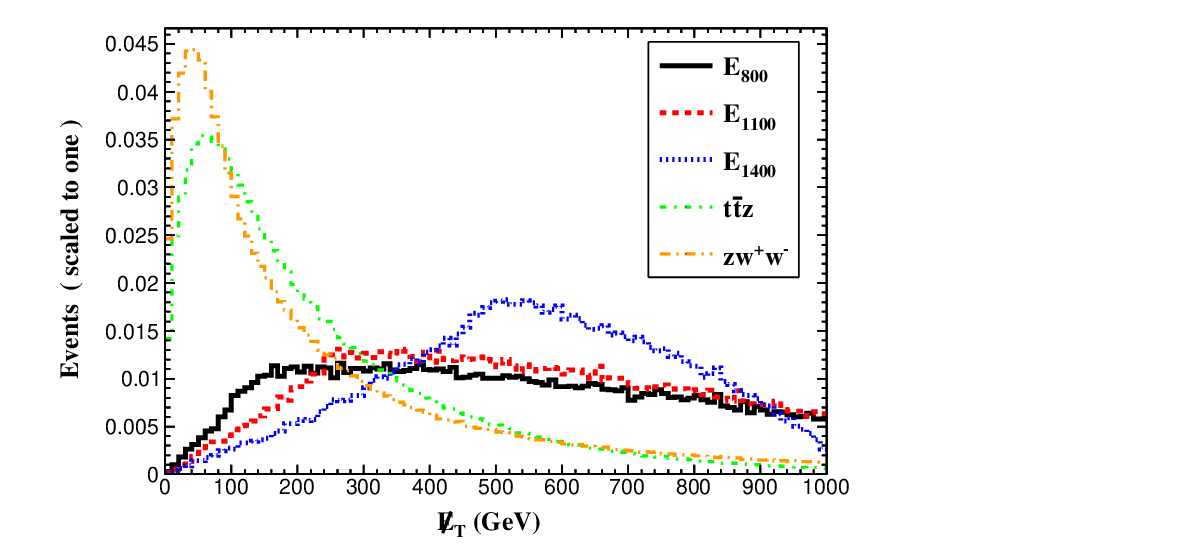}}
\centerline{\hspace{2.0cm}\epsfxsize=9cm\epsffile{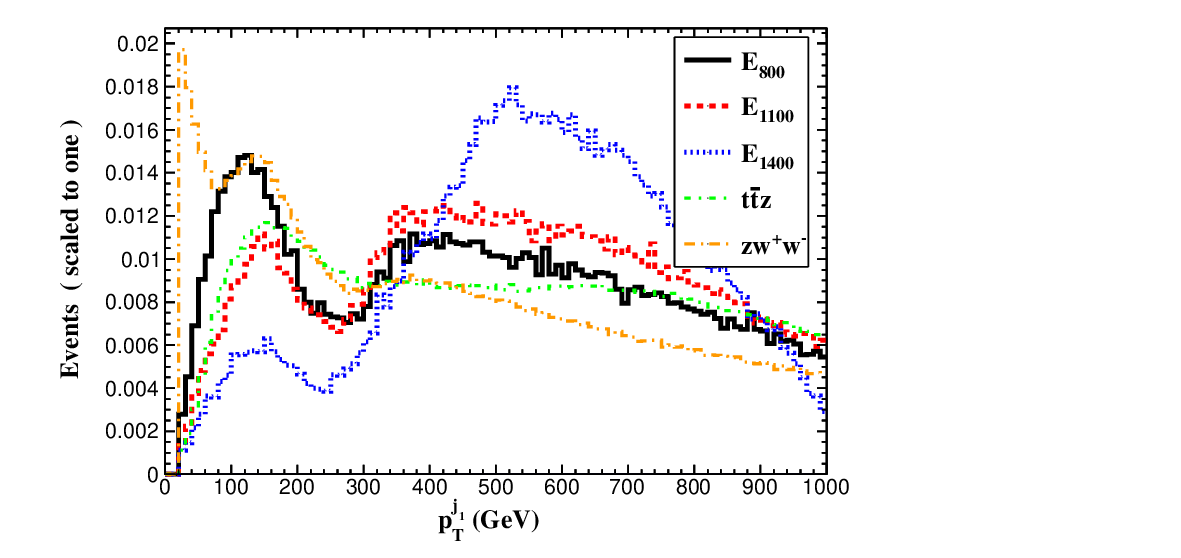}
\hspace{-2.0cm}\epsfxsize=9cm\epsffile{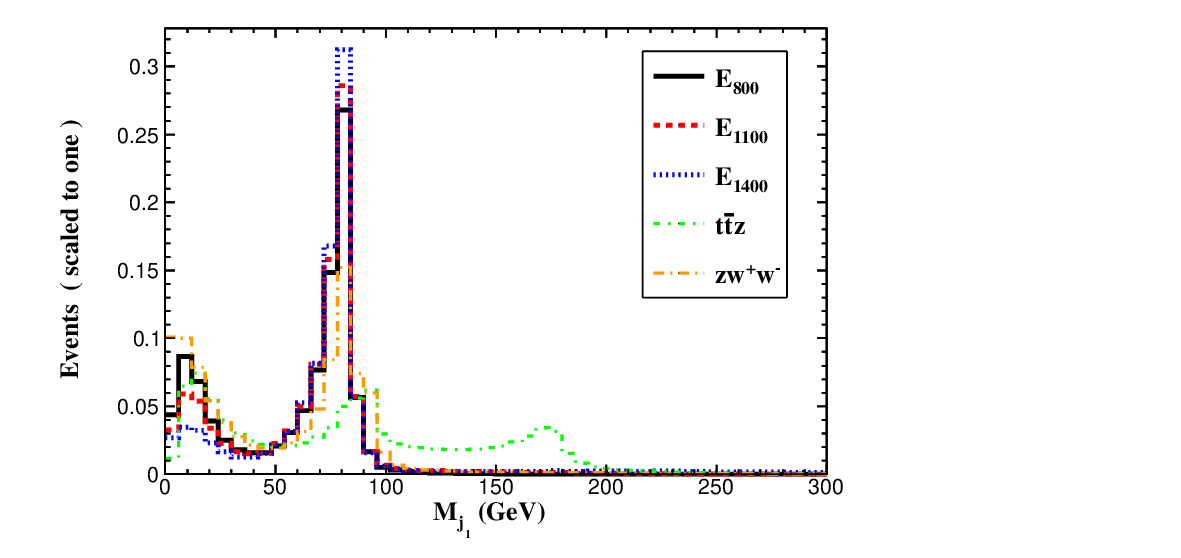}}
\caption{Same as Figure~\ref{fig:distribution2-1}, but for $\sqrt{s} = 3$ TeV.  }
\label{fig:distribution2-3}
\end{center}
\end{figure*}

Based on the kinematic differences between signal and background processes, we show in Figures~\ref{fig:distribution2-1}--\ref{fig:distribution2-3} the differential distributions for the signals and relevant SM backgrounds.
These observables offer distinctive features for signal identification and background suppression.

The event selection is performed through three sequential cut stages.
\begin{itemize}
\item {Cut 1: lepton selection}
\begin{itemize}
\item Exactly three isolated leptons ($e$ or $\mu$) are required: $N_{\ell}=3$.
\item The leading lepton must satisfy the transverse momentum thresholds of $p_T^{\ell_1} > \SI{120}{GeV}$ at $\sqrt{s} = \SI{1}{TeV}$, $> \SI{150}{GeV}$ at $\SI{1.5}{TeV}$ and $> \SI{250}{GeV}$ at $\SI{3}{TeV}$.
\item At least one opposite-sign same-flavor lepton pair must be consistent with a $Z$ boson decay: $|M_{\ell\ell} - m_Z| < \SI{10}{GeV}$.
\end{itemize}
\item {Cut 2: missing transverse energy and jet requirements}
\begin{itemize}
\item The missing transverse energy must satisfy $\slashed{E}_T > \SI{80}{GeV}$ at $\sqrt{s} = \SI{1}{TeV}$, $> \SI{100}{GeV}$ at $\SI{1.5}{TeV}$ and $> \SI{300}{GeV}$ at $\SI{3}{TeV}$.
\item For $\sqrt{s} = \SI{1}{TeV}$ and $\SI{1.5}{TeV}$: the dijet transverse mass must exceed $\SI{250}{GeV}$ and $\SI{350}{GeV}$, respectively.
\item For $\sqrt{s} = \SI{3}{TeV}$: at least one fat jet is required with $p_T^{j_1} > \SI{300}{GeV}$ and invariant mass $\SI{60}{GeV} < M_{j_1} < \SI{110}{GeV}$.
\end{itemize}
\item {Cut 3: trilepton invariant mass selection}
\begin{itemize}
\item The minimum requirement for the trilepton invariant mass is: $M_{\ell_{1}\ell_{2}\ell_{3}} > \SI{300}{GeV}$ at $\sqrt{s} = \SI{1}{TeV}$, $M_{\ell_{1}\ell_{2}\ell_{3}} > \SI{450}{GeV}$ at $\sqrt{s} = \SI{1.5}{TeV}$ and $M_{\ell_{1}\ell_{2}\ell_{3}} > \SI{600}{GeV}$ at $\sqrt{s} = \SI{3}{TeV}$.
\end{itemize}
\end{itemize}

\begin{table}[htb]
\centering
\small
\caption{Cross sections (in fb) for our three signal benchmarks (with different $m_E$) and SM backgrounds in Case 2 at $\sqrt{s} = 1$~TeV. Values in parentheses show the selection efficiency after the corresponding cut is applied.\label{cutflow2-1}}
\vspace{0.1cm}
\begin{tabular}{@{}l ccccl cc@{}}
\toprule[1.5pt]
\multirow{2}{*}{Cuts} & \multicolumn{3}{c}{Signal ($m_{E}$ in GeV)} & \phantom{a} & \multicolumn{2}{c}{Background} \\
\cmidrule(lr){2-4} \cmidrule(lr){6-7}
& 350 &400 & 450 & & $t\bar{t}Z$ & $ZW^{+}W^{-}$ \\
\midrule[0.8pt]
Basic & 1.82 & 1.58 &1.19 & & 0.98 & 3.63\\
Cut 1 & 0.55 (30\%) & 0.48 (30\%) &0.36 (30\%) & & 0.0048 (0.17\%) & 0.036 (0.35\%) \\
Cut 2 & 0.20 (11\%) & 0.22 (13.6\%) &0.20 (16.7\%) & & 3.3e-4 (0.012\%) & 0.0034 (0.034\%) \\
Cut 3 & 0.20 (11\%) & 0.22 (13.6\%) &0.20 (16.7\%) & & 1.2e-4 (0.0046\%) & 0.002 (0.02\%) \\
\bottomrule[1.5pt]
\end{tabular}
\end{table}

\begin{table}[htb]
\centering
\small
\caption{Same as Table~\ref{cutflow2-1} but for $\sqrt{s} = 1.5$~TeV.\label{cutflow2-2}}
\vspace{0.1cm}
\begin{tabular}{@{}l ccccl cc@{}}
\toprule[1.5pt]
\multirow{2}{*}{Cuts} & \multicolumn{3}{c}{Signal ($m_{E}$ in GeV)} & \phantom{a} & \multicolumn{2}{c}{Background} \\
\cmidrule(lr){2-4} \cmidrule(lr){6-7}
& 500 & 600 &700 & & $t\bar{t}Z$ & $ZW^{+}W^{-}$ \\
\midrule[0.8pt]
Basic & 0.78 & 0.66 &0.42& & 0.63 & 3.58 \\
Cut 1 & 0.24 (30\%) & 0.20 (30\%) &0.14 (33\%)& & 0.0029 (0.15\%) & 0.035 (0.34\%) \\
Cut 2 & 0.066 (8.3\%) & 0.068 (10\%)&0.054 (12.5\%) & & 6.4e-5 (0.0034\%) & 0.0034 (0.017\%) \\
Cut 3 & 0.066 (8.3\%) & 0.067 (10\%)&0.054  (12.5\%)& & 2.2e-5 (0.0012\%) & 0.002 (0.01\%) \\
\bottomrule[1.5pt]
\end{tabular}
\end{table}
\begin{table}[htb]
\centering
\small
\caption{Same as Table~\ref{cutflow2-1} but for $\sqrt{s} = 3$~TeV.\label{cutflow2-3}}
\vspace{0.1cm}
\begin{tabular}{@{}l ccccl cc@{}}
\toprule[1.5pt]
\multirow{2}{*}{Cuts} & \multicolumn{3}{c}{Signal ($m_{E}$ in GeV)} & \phantom{a} & \multicolumn{2}{c}{Background} \\
\cmidrule(lr){2-4} \cmidrule(lr){6-7}
& 800 & 1100 &1400 & & $t\bar{t}Z$ & $ZW^{+}W^{-}$ \\
\midrule[0.8pt]
Basic & 0.19 & 0.17 &0.10& & 0.28 & 2.17 \\
Cut 1 & 0.055 (28\%) & 0.048 (28\%) & 0.032 (30\%) & & 0.0015 (0.17\%) & 0.02 (0.3\%) \\
Cut 2 & 0.023 (12\%) & 0.025 (14\%) & 0.021 (20\%) & & 8.7e-5 (0.01\%) & 0.0022 (0.032\%) \\
Cut 3 & 0.023 (12\%) & 0.025 (14\%) & 0.021 (20\%) & & 5.0e-5 (0.0057\%) & 0.0013 (0.019\%) \\
\bottomrule[1.5pt]
\end{tabular}
\end{table}

The cutflow data in Tables~\ref{cutflow2-1}--\ref{cutflow2-3} detail the evolution of cross sections and cumulative selection efficiencies for both the usual three signal benchmarks  and the dominant SM backgrounds ($t\bar{t}Z$ and $ZW^{+}W^{-}$) in Case 2 at c.m. energies of 1, 1.5 and 3 TeV, revealing that the signal efficiencies remain relatively stable (approximately 8\%-20\% after full selection) and are generally higher for heavier masses. In contrast, the background efficiencies are dramatically suppressed by several orders of magnitude, falling to as low as $10^{-5}$ for $t\bar{t}Z$ and $10^{-4}$ for $ZW^{+}W^{-}$ after the final selection, thereby underscoring the exceptional effectiveness of the sequential cuts in enhancing the signal-to-background ratio.

\subsection{Statistical analysis}
The discovery ($\mathcal{Z}_{\text{disc}}$) and exclusion ($\mathcal{Z}_{\text{excl}}$) significances are calculated following the method in Ref.~\cite{Cowan:2010js}:
\beq
\begin{aligned}
\mathcal{Z}_\text{disc} &= \sqrt{2\left[(s+b)\ln\left(1+\frac{s}{b}\right)-s\right]}, \\
\mathcal{Z}_\text{excl} &= \sqrt{2\left[s - b\ln\left(1+\frac{s}{b}\right)\right]},
\end{aligned}
\eeq
where $s$ and $b$ represent the expected numbers of signal and background events, respectively, for a given integrated luminosity.

\begin{figure}[htb]
\begin{center}
\centerline{\hspace{1.0cm}\epsfxsize=6.5cm\epsffile{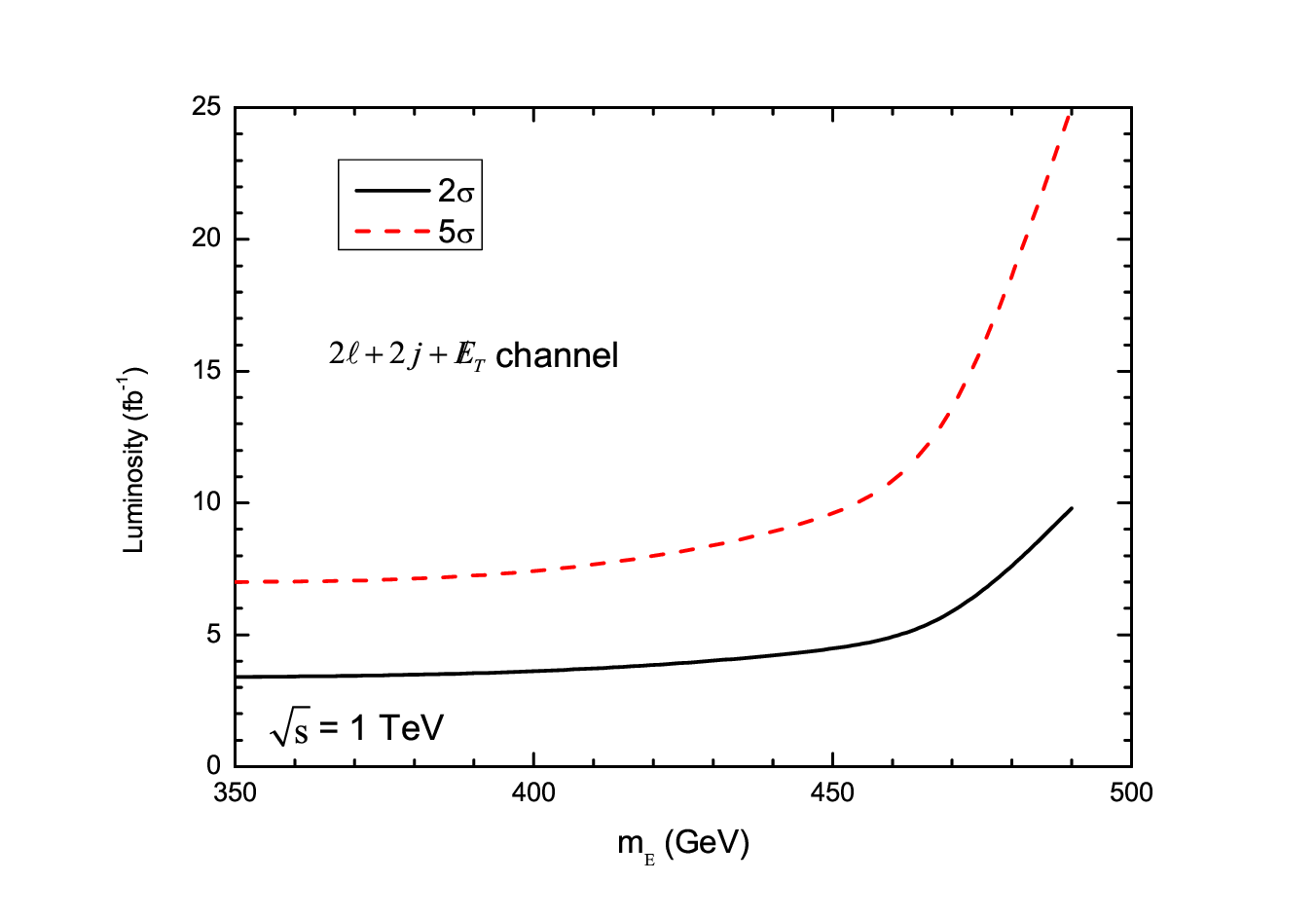}
\hspace{-1cm}\epsfxsize=6.5cm\epsffile{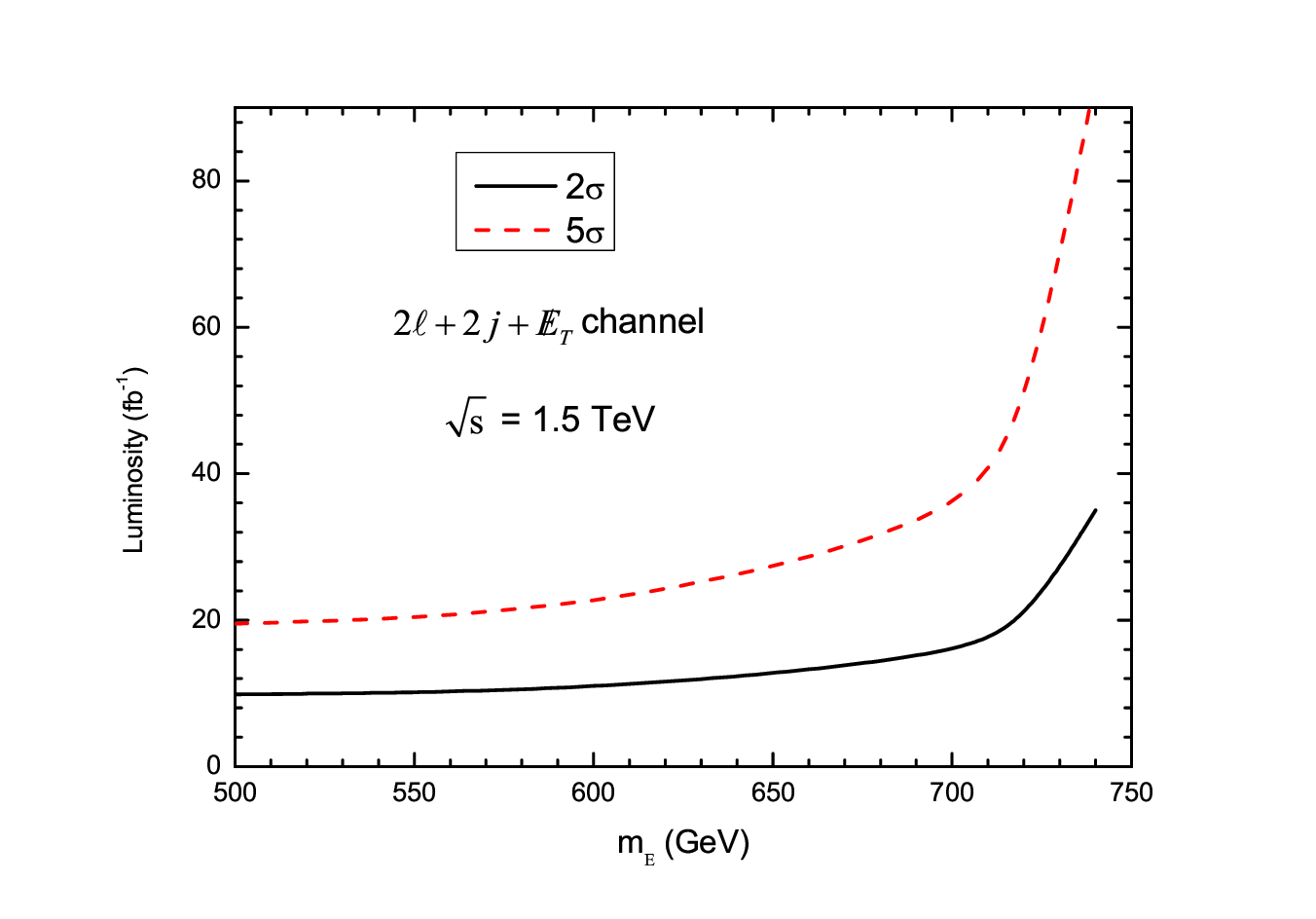}\hspace{-1cm}\epsfxsize=6.5cm\epsffile{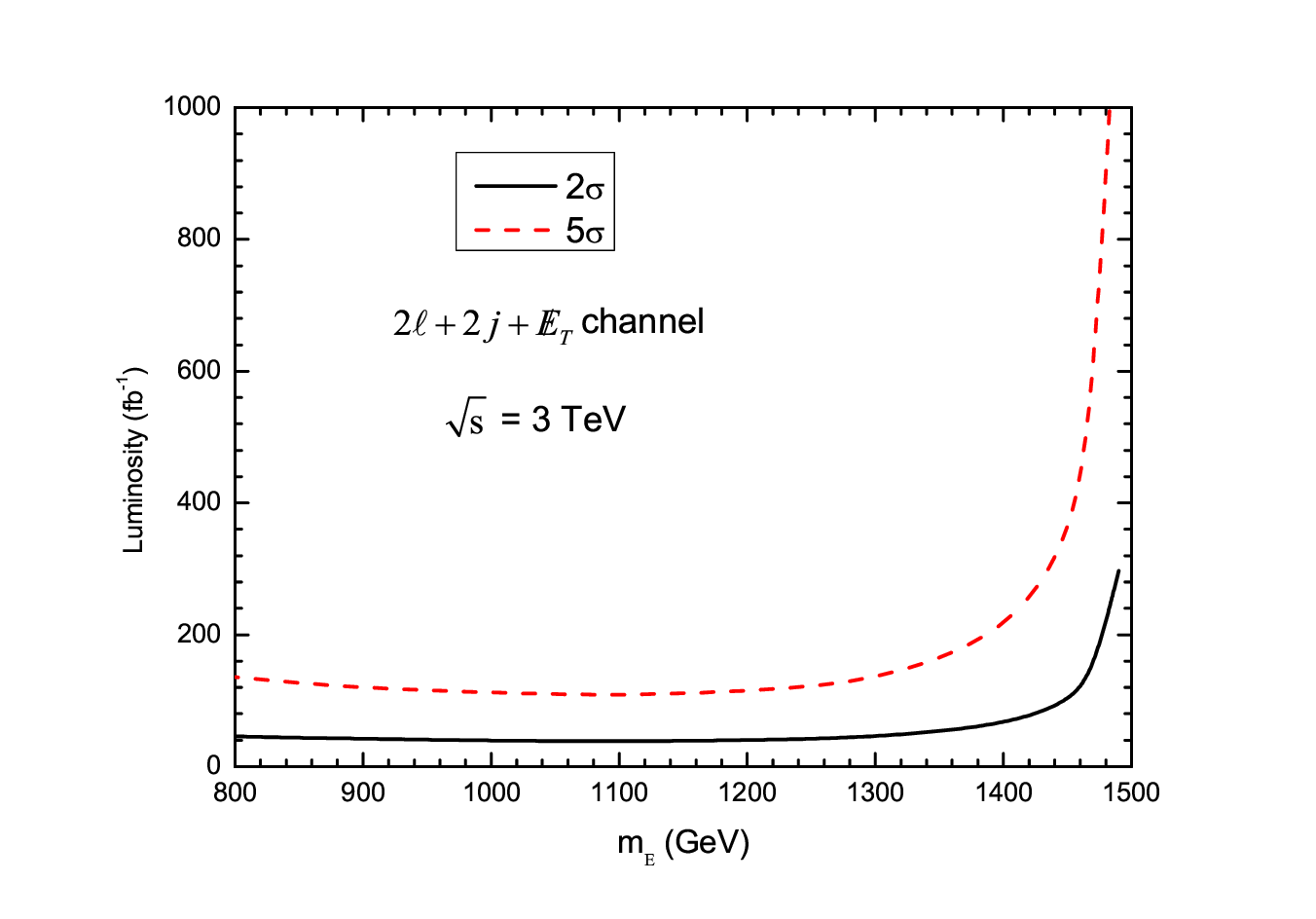}}
\centerline{\hspace{1.0cm}\epsfxsize=6.5cm\epsffile{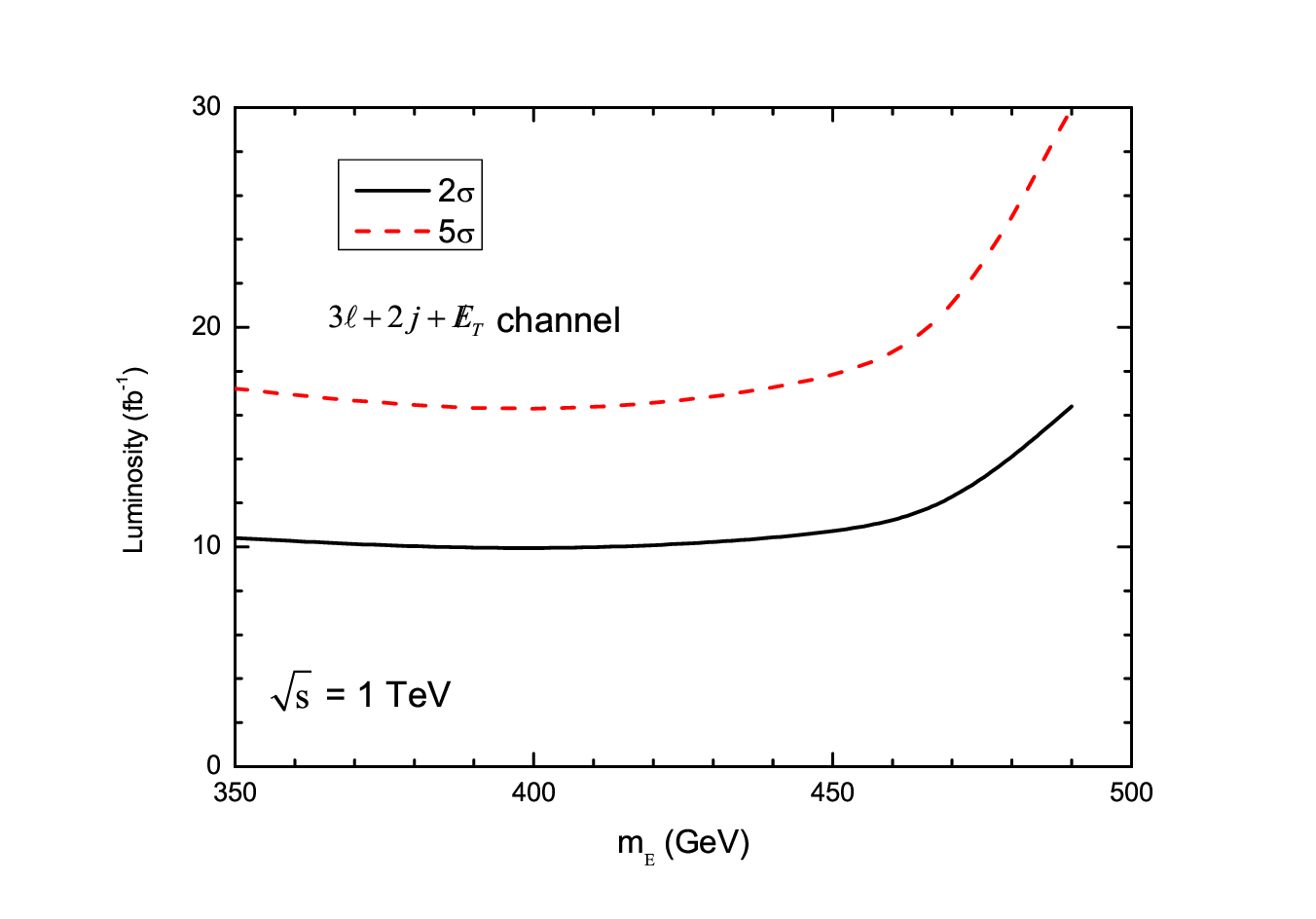}
\hspace{-1cm}\epsfxsize=6.5cm\epsffile{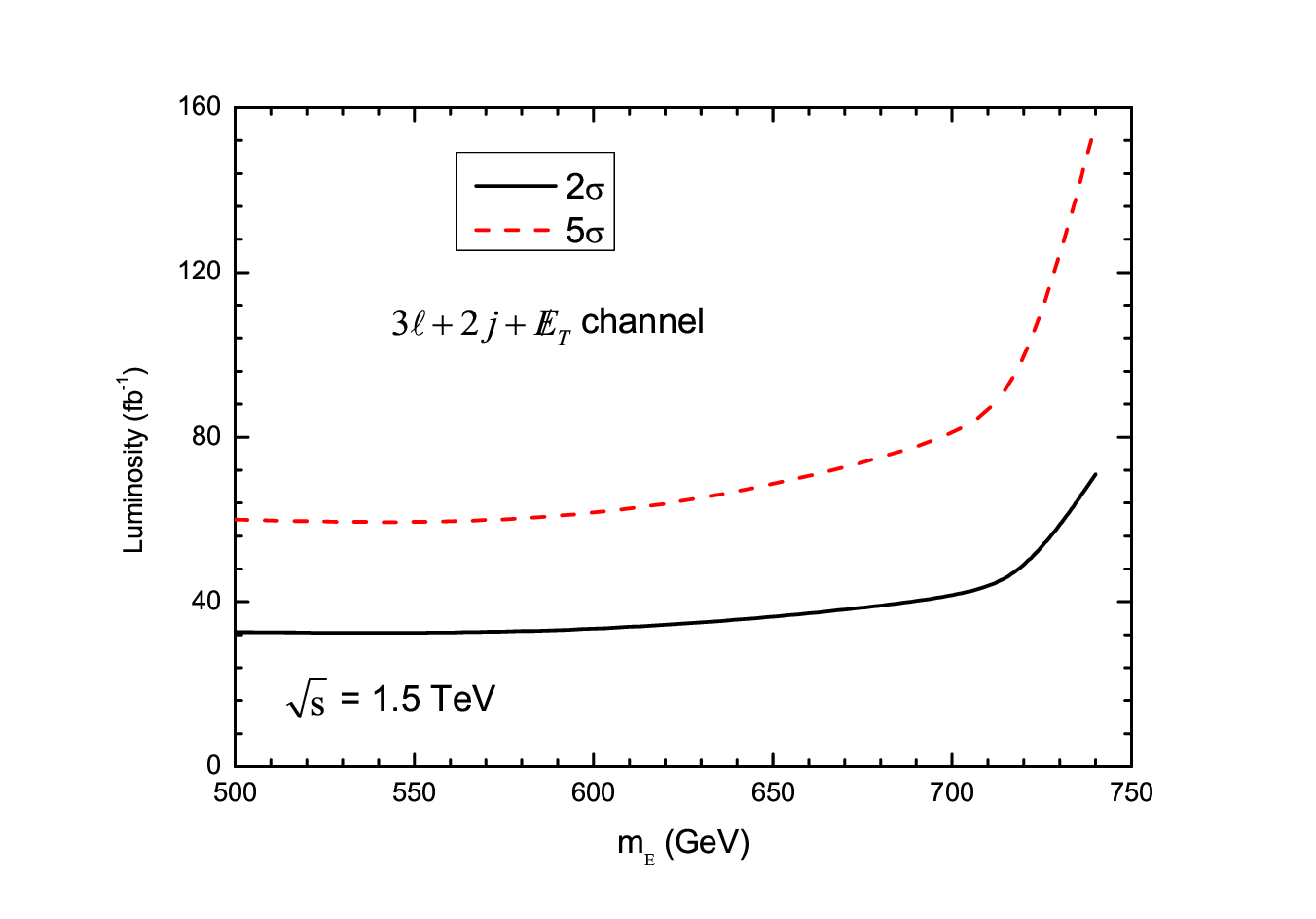}\hspace{-1cm}\epsfxsize=6.5cm\epsffile{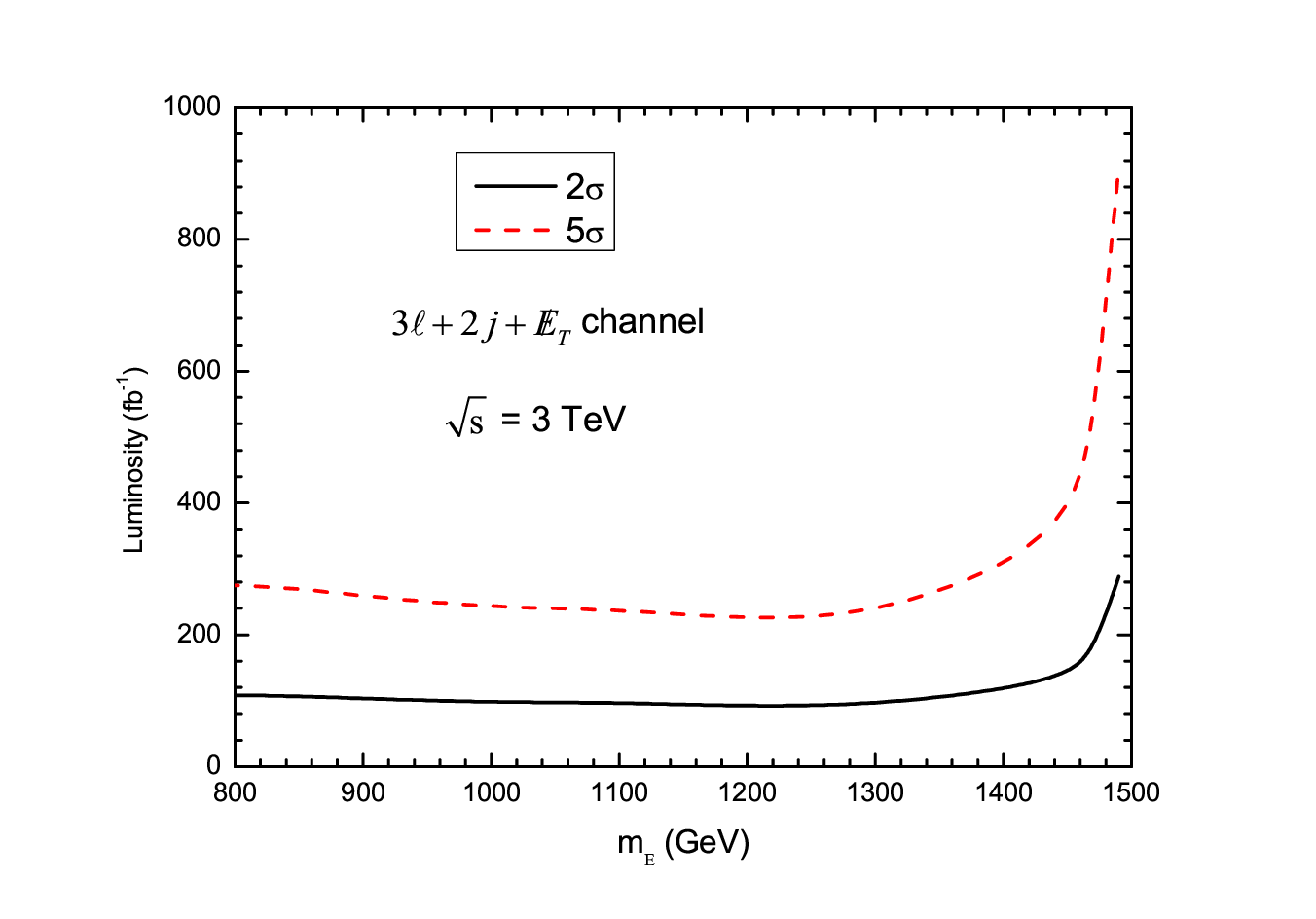}}
\caption{$2\sigma$ exclusion limits (solid curves) and $5\sigma$ discovery prospects (dashed curves) are shown as a function of the heavy VLL mass $m_E$ for Case 1 (top) and Case 2 (bottom), at three c.m. energies: 1 TeV (left), 1.5 TeV (middle) and 3 TeV (right). }
\label{fig-combined-sensitivity}
\end{center}
\end{figure}

The combined sensitivity for both $2\sigma$ exclusion and $5\sigma$ discovery is presented in Figure~\ref{fig-combined-sensitivity}, which displays the required integrated luminosity as a function of $m_E$ for both final states considered at $\sqrt{s} = \SI{1}{TeV}$, \SI{1.5}{TeV} and \SI{3}{TeV}. The contours clearly demonstrate that the integrated luminosity requirement grows gradually with increasing $m_E$ and exhibits a steep rise near the kinematic threshold, mirroring the rapid decline in signal cross section as $m_E$ approaches the production limit. Case~2 consistently demands higher integrated luminosity compared to Case~1 for both exclusion and discovery sensitivities. The $2\ell+2j+\slashed{E}_T$ final state proves to be especially promising for probing singlet vector-like leptons of electron flavor at future $e^+e^-$ colliders. In particular, at $\sqrt{s}=\SI{1}{TeV}$, masses up to $m_E \lesssim \SI{490}{GeV}$ can be explored with $\SI{25}{fb^{-1}}$, at $\sqrt{s}=\SI{1.5}{TeV}$, the mass reach extends to $m_E \lesssim \SI{740}{GeV}$ with $\SI{90}{fb^{-1}}$; while at $\sqrt{s}=\SI{3}{TeV}$ masses up to $m_E \lesssim \SI{1440}{GeV}$ become accessible with $\SI{1000}{fb^{-1}}$. The $3\ell+2j+\slashed{E}_T$ signature can instead be used for signal consolidation purposes.
\section{Summary}
\label{sec:conclusion}
The clean experimental environment of high energy $e^+e^-$ colliders, together with beam polarization capabilities and well-defined initial states, offers distinct advantages in searches for VLLs over hadronic environments. In this work, we have performed a comprehensive analysis of the pair production and decay patterns of weak-isosinglet VLL of electron flavor at future $e^+e^-$ colliders. Using detailed MC simulations at c.m. energies of $\sqrt{s} = \SI{1}{TeV}$, $\SI{1.5}{TeV}$ and $\SI{3}{TeV}$, including beam polarization effects, we have developed optimized search strategies targeting both $2\ell+2j+\slashed{E}_T$ and $3\ell+2j+\slashed{E}_T$ final states.

Upon adopting a variety of $E^\pm$ masses within the kinematic reach of the ILC and CLIC prototypes considered here, each with
varying luminosity values,
our results demonstrate that well-designed selection criteria can suppress SM backgrounds efficiently while retaining a high signal acceptance. The first channel is especially powerful for singlet VLL searches, owing to its sizable production cross section and favorable signal-to-background ratio.  The projected discovery reaches at $5\sigma$ significance extend to $m_E \lesssim \SI{490}{GeV}$ at $\sqrt{s} = \SI{1}{TeV}$ with $\SI{25}{fb^{-1}}$, $m_E \lesssim \SI{740}{GeV}$ at $\sqrt{s} = \SI{1.5}{TeV}$ with $\SI{90}{fb^{-1}}$ and $m_E \lesssim \SI{1440}{GeV}$ at $\sqrt{s} = \SI{3}{TeV}$ with $\SI{1000}{fb^{-1}}$, substantially exceeding the current mass limits from hadron colliders. The second channel  is less sensitive in comparison but can be used for signal confirmation.

\begin{acknowledgments}
\vspace*{-0.5truecm}
Y-BL  is supported by the Natural Science Foundation of Henan Province (Grant No. 252300421988).
SM is supported  through the NExT Institute and STFC CG ST/X000583/1.
\end{acknowledgments}


\clearpage

\begin{thebibliography}{99}
\bibitem{Arkani-Hamed:2012dcq}
N.~Arkani-Hamed, K.~Blum, R.~T.~D'Agnolo and J.~Fan,
\href{doi:10.1007/JHEP01(2013)149}{JHEP \textbf{01}, 149 (2013)}.
[arXiv:1207.4482 [hep-ph]].


\bibitem{Schwaller:2013hqa}
P.~Schwaller, T.~M.~P.~Tait and R.~Vega-Morales,
\href{doi:10.1103/PhysRevD.88.035001}{Phys.\ Rev.\ D \textbf{88}, no.3, 035001 (2013)}.
[arXiv:1305.1108 [hep-ph]].

\bibitem{Halverson:2014nwa}
J.~Halverson, N.~Orlofsky and A.~Pierce,
\href{doi:10.1103/PhysRevD.90.015002}{Phys.\ Rev.\ D \textbf{90}, no.1, 015002 (2014)}.
[arXiv:1403.1592 [hep-ph]].

\bibitem{Bahrami:2016has}
S.~Bahrami, M.~Frank, D.~K.~Ghosh, N.~Ghosh and I.~Saha,
\href{doi:10.1103/PhysRevD.95.095024}{Phys.\ Rev.\ D \textbf{95}, no.9, 095024 (2017)}.
[arXiv:1612.06334 [hep-ph]].

\bibitem{Bhattacharya:2018fus}
S.~Bhattacharya, P.~Ghosh, N.~Sahoo and N.~Sahu,
\href{doi:10.3389/fphy.2019.00080}{Front.\ in Phys.\ \textbf{7}, 80 (2019)}.
[arXiv:1812.06505 [hep-ph]].

\bibitem{Xiao:2014kba}
M.~L.~Xiao and J.~H.~Yu,
\href{doi:10.1103/PhysRevD.90.014007}{Phys.\ Rev.\ D \textbf{90}, no.1, 014007 (2014)}.
[arXiv:1404.0681 [hep-ph]].

\bibitem{Cingiloglu:2024vdh}
K.~Y.~Cingiloglu and M.~Frank,
\href{doi:10.1103/PhysRevD.111.016025}{Phys.\ Rev.\ D \textbf{111}, no.1, 016025 (2025)}.
[arXiv:2408.10898 [hep-ph]].

\bibitem{Hiller:2019mou}
G.~Hiller, C.~Hormigos-Feliu, D.~F.~Litim and T.~Steudtner,
\href{doi:10.1103/PhysRevD.102.071901}{Phys.\ Rev.\ D \textbf{102}, no.7, 071901 (2020)}.
[arXiv:1910.14062 [hep-ph]].

\bibitem{DeJesus:2020yqx}
A.~S.~De Jesus, S.~Kovalenko, F.~S.~Queiroz, C.~Siqueira and K.~Sinha,
\href{doi:10.1103/PhysRevD.102.035004}{Phys.\ Rev.\ D \textbf{102}, no.3, 035004 (2020)}.
[arXiv:2004.01200 [hep-ph]].

\bibitem{Frank:2020smf}
M.~Frank and I.~Saha,
\href{doi:10.1103/PhysRevD.102.115034}{Phys.\ Rev.\ D \textbf{102}, no.11, 115034 (2020)}.
[arXiv:2008.11909 [hep-ph]].

\bibitem{Dermisek:2021ajd}
R.~Dermisek, K.~Hermanek and N.~McGinnis,
\href{doi:10.1103/PhysRevD.104.055033}{Phys.\ Rev.\ D \textbf{104}, no.5, 055033 (2021)}.
[arXiv:2103.05645 [hep-ph]].

\bibitem{Brune:2022rlo}
T.~Brune, T.~W.~Kephart and H.~P\"as,
\href{doi:10.1140/epjc/s10052-024-13617-5}{Eur.\ Phys.\ J.\ C \textbf{84}, no.12, 1254 (2024)}.
[arXiv:2205.05566 [hep-ph]].

\bibitem{Guedes:2022cfy}
G.~Guedes and P.~Olgoso,
\href{doi:10.1007/JHEP09(2022)181}{JHEP \textbf{09}, 181 (2022)}.
[arXiv:2205.04480 [hep-ph]].

\bibitem{He:2022zjz}
S.~P.~He,
\href{doi:10.1088/1674-1137/ac9e4c}{Chin.\ Phys.\ C \textbf{47}, no.4, 043102 (2023)}.
[arXiv:2205.02088 [hep-ph]].



\bibitem{Kawamura:2022fhm}
J.~Kawamura and S.~Raby,
\href{doi:10.1103/PhysRevD.106.035009}{Phys.\ Rev.\ D \textbf{106}, no.3, 035009 (2022)}.
[arXiv:2205.10480 [hep-ph]].





\bibitem{Huang:2012kz}
G.~Y.~Huang, K.~Kong and S.~C.~Park,
\href{doi:10.1007/JHEP06(2012)099}{JHEP \textbf{06}, 099 (2012)}.
[arXiv:1204.0522 [hep-ph]].

\bibitem{Kong:2010qd}
K.~Kong, S.~C.~Park and T.~G.~Rizzo,
\href{doi:10.1007/JHEP07(2010)059}{JHEP \textbf{07}, 059 (2010)}.
[arXiv:1004.4635 [hep-ph]].

\bibitem{Graham:2009gy}
P.~W.~Graham, A.~Ismail, S.~Rajendran and P.~Saraswat,
\href{doi:10.1103/PhysRevD.81.055016}{Phys.\ Rev.\ D \textbf{81}, 055016 (2010)}.
[arXiv:0910.3020 [hep-ph]].

\bibitem{Endo:2011xq}
M.~Endo, K.~Hamaguchi, S.~Iwamoto and N.~Yokozaki,
\href{doi:10.1103/PhysRevD.85.095012}{Phys.\ Rev.\ D \textbf{85}, 095012 (2012)}.
[arXiv:1112.5653 [hep-ph]].

\bibitem{Martin:2012dg}
S.~P.~Martin and J.~D.~Wells,
\href{doi:10.1103/PhysRevD.86.035017}{Phys.\ Rev.\ D \textbf{86}, 035017 (2012)}.
[arXiv:1206.2956 [hep-ph]].

\bibitem{Endo:2012cc}
M.~Endo, K.~Hamaguchi, K.~Ishikawa, S.~Iwamoto and N.~Yokozaki,
\href{doi:10.1007/JHEP01(2013)181}{JHEP \textbf{01}, 181 (2013)}.
[arXiv:1212.3935 [hep-ph]].

\bibitem{Fischler:2013tva}
W.~Fischler and W.~Tangarife,
\href{doi:10.1007/JHEP05(2014)151}{JHEP \textbf{05}, 151 (2014)}.
[arXiv:1310.6369 [hep-ph]].

\bibitem{DeCurtis:2018iqd}
S.~De Curtis, L.~Delle Rose, S.~Moretti and K.~Yagyu,
\href{doi:10.1016/j.physletb.2018.09.042}{Phys. Lett. B \textbf{786}, 189 (2018)}.
[arXiv:1803.01865 [hep-ph]].

\bibitem{He:1999vp}
H.~J.~He, T.~M.~P.~Tait and C.~P.~Yuan,
\href{doi:10.1103/PhysRevD.62.011702}{Phys.\ Rev.\ D \textbf{62}, 011702(R) (2000)}.
[arXiv:hep-ph/9911266 [hep-ph]].

\bibitem{Wang:2013jwa}
X.~F.~Wang, C.~Du and H.~J.~He,
\href{doi:10.1016/j.physletb.2013.05.015}{Phys.\ Lett.\ B \textbf{723}, 314-323 (2013)}.
[arXiv:1304.2257 [hep-ph]].

\bibitem{He:2001fz}
H.~J.~He, C.~T.~Hill and T.~M.~P.~Tait,
\href{doi:10.1103/PhysRevD.65.055006}{Phys.\ Rev.\ D \textbf{65}, 055006 (2002)}.
[arXiv:hep-ph/0108041 [hep-ph]].

\bibitem{He:2014ora}
H.~J.~He and Z.~Z.~Xianyu,
\href{doi:10.1088/1475-7516/2014/10/019}{JCAP\ \textbf{10}, 019 (2014)}.
[arXiv:1405.7331 [hep-ph]].

\bibitem{delAguila:2008pw}
F.~del Aguila, J.~de Blas and M.~Perez-Victoria,
\href{doi:10.1103/PhysRevD.78.013010}{Phys.\ Rev.\ D \textbf{78}, 013010 (2008)}.
[arXiv:0803.4008 [hep-ph]].

\bibitem{Ishiwata:2013gma}
K.~Ishiwata and M.~B.~Wise,
\href{doi:10.1103/PhysRevD.88.055009}{Phys.\ Rev.\ D \textbf{88}, 055009 (2013)}.
[arXiv:1307.1112 [hep-ph]].

\bibitem{Kearney:2012zi}
J.~Kearney, A.~Pierce and N.~Weiner,
\href{doi:10.1103/PhysRevD.86.113005}{Phys.\ Rev.\ D \textbf{86}, 113005 (2012)}.
[arXiv:1207.7062 [hep-ph]].

\bibitem{Altmannshofer:2013zba}
W.~Altmannshofer, M.~Bauer and M.~Carena,
\href{doi:10.1007/JHEP01(2014)060}{JHEP \textbf{01}, 060 (2014)}.
[arXiv:1308.1987 [hep-ph]].

\bibitem{Falkowski:2013jya}
A.~Falkowski, D.~M.~Straub and A.~Vicente,
\href{doi:10.1007/JHEP05(2014)092}{JHEP \textbf{05}, 092 (2014)}.
[arXiv:1312.5329 [hep-ph]].


\bibitem{Ellis:2014dza}
S.~A.~R.~Ellis, R.~M.~Godbole, S.~Gopalakrishna and J.~D.~Wells,
\href{doi:10.1007/JHEP09(2014)130}{JHEP \textbf{09}, 130 (2014)}.
[arXiv:1404.4398 [hep-ph]].

\bibitem{Ishiwata:2015cga}
K.~Ishiwata, Z.~Ligeti and M.~B.~Wise,
\href{doi:10.1007/JHEP10(2015)027}{JHEP \textbf{10}, 027 (2015)}.
[arXiv:1506.03484 [hep-ph]].

\bibitem{Dermisek:2014cia}
R.~Dermisek, A.~Raval and S.~Shin,
\href{doi:10.1103/PhysRevD.90.034023}{Phys.\ Rev.\ D \textbf{90}, 034023 (2014)}.
[arXiv:1406.7018 [hep-ph]].

\bibitem{Crivellin:2020ebi}
A.~Crivellin, F.~Kirk, C.~A.~Manzari and M.~Montull,
\href{doi:10.1007/JHEP12(2020)166}{JHEP \textbf{12}, 166 (2020)}.
[arXiv:2008.01113 [hep-ph]].

\bibitem{Endo:2020tkb}
M.~Endo and S.~Mishima,
\href{doi:10.1007/JHEP08(2020)004}{JHEP \textbf{08}, 004 (2020)}.
[arXiv:2005.03933 [hep-ph]].

\bibitem{Chakrabarty:2020jro}
N.~Chakrabarty,
\href{doi:10.1140/epjp/s13360-021-02168-3}{Eur.\ Phys.\ J.\ Plus \textbf{136}, 1183 (2021)}.
[arXiv:2010.05215 [hep-ph]].

\bibitem{Guedes:2021oqx}
G.~Guedes and J.~Santiago,
\href{doi:10.1007/JHEP01(2022)111}{JHEP \textbf{01}, 111 (2022)}.
[arXiv:2107.03429 [hep-ph]].

\bibitem{Raju:2022zlv}
M.~Raju, A.~Mukherjee and J.~P.~Saha,
\href{doi:10.1140/epjc/s10052-023-11595-8}{Eur.\ Phys.\ J.\ C \textbf{83}, 429 (2023)}.
[arXiv:2207.02825 [hep-ph]].

\bibitem{Li:2023mrw}
X.~Q.~Li, Z.~J.~Xie, Y.~D.~Yang and X.~B.~Yuan,
\href{doi:10.1016/j.nuclphysb.2024.116646}{Nucl.\ Phys.\ B \textbf{1006}, 116646 (2024)}.
[arXiv:2307.05290 [hep-ph]].

\bibitem{Dermisek:2014qca}
R.~Dermisek, J.~P.~Hall, E.~Lunghi and S.~Shin,
\href{doi:10.1007/JHEP12(2014)013}{JHEP \textbf{12}, 013 (2014)}.
[arXiv:1408.3123 [hep-ph]].

\bibitem{Dermisek:2015oja}
R.~Dermisek, E.~Lunghi and S.~Shin,
\href{doi:10.1007/JHEP02(2016)119}{JHEP \textbf{02}, 119 (2016)}.
[arXiv:1509.04292 [hep-ph]].

\bibitem{Kumar:2015tna}
N.~Kumar and S.~P.~Martin,
\href{doi:10.1103/PhysRevD.92.115018}{Phys.\ Rev.\ D \textbf{92}, 115018 (2015)}.
[arXiv:1510.03456 [hep-ph]].

\bibitem{Dermisek:2015hue}
R.~Dermisek, E.~Lunghi and S.~Shin,
\href{doi:10.1007/JHEP05(2016)148}{JHEP \textbf{05}, 148 (2016)}.
[arXiv:1512.07837 [hep-ph]].

\bibitem{Chen:2016lsr}
C.~H.~Chen and T.~Nomura,
\href{doi:10.1140/epjc/s10052-016-4197-3}{Eur.\ Phys.\ J.\ C \textbf{76}, 353 (2016)}.
[arXiv:1602.07519 [hep-ph]].

\bibitem{Kawamura:2019rth}
J.~Kawamura, S.~Raby and A.~Trautner,
\href{doi:10.1103/PhysRevD.100.055030}{Phys.\ Rev.\ D \textbf{100}, 055030 (2019)}.
[arXiv:1906.11297 [hep-ph]].

\bibitem{Freitas:2020ttd}
F.~F.~Freitas, J.~Gon\c{c}alves, A.~P.~Morais and R.~Pasechnik,
\href{doi:10.1007/JHEP01(2021)076}{JHEP \textbf{01}, 076 (2021)}.
[arXiv:2010.01307 [hep-ph]].

\bibitem{OsmanAcar:2021plv}
A.~Osman Acar, O.~E.~Delialioglu and S.~Sultansoy,
[arXiv:2103.08222 [hep-ph]].

\bibitem{Kawamura:2021ygg}
J.~Kawamura and S.~Raby,
\href{doi:10.1103/PhysRevD.104.035007}{Phys.\ Rev.\ D \textbf{104}, 035007 (2021)}.
[arXiv:2104.04461 [hep-ph]].

\bibitem{Bonilla:2021ize}
C.~Bonilla, A.~E.~C\'arcamo Hern\'andez, J.~Gon\c{c}alves, F.~F.~Freitas, A.~P.~Morais and R.~Pasechnik,
\href{doi:10.1007/JHEP01(2022)154}{JHEP \textbf{01}, 154 (2022)}.
[arXiv:2107.14165 [hep-ph]].

\bibitem{Baspehlivan:2022qet}
F.~Baspehlivan, B.~Dagli, O.~E.~Delialioglu and S.~Sultansoy,
[arXiv:2201.08251 [hep-ph]].

\bibitem{Cao:2023smj}
Q.~H.~Cao, J.~Guo, J.~Liu, Y.~Luo and X.~P.~Wang,
\href{doi:10.1103/PhysRevD.110.015029}{Phys.\ Rev.\ D \textbf{110}, 015029 (2024)}.
[arXiv:2311.12934 [hep-ph]].

\bibitem{Bernreuther:2023uxh}
E.~Bernreuther and B.~A.~Dobrescu,
\href{doi:10.1007/JHEP07(2023)079}{JHEP \textbf{07}, 079 (2023)}.
[arXiv:2304.08509 [hep-ph]].

\bibitem{Kawamura:2023zuo}
J.~Kawamura and S.~Shin,
\href{doi:10.1007/JHEP11(2023)025}{JHEP \textbf{11}, 025 (2023)}.
[arXiv:2308.07814 [hep-ph]].


\bibitem{Bigaran:2023ris}
I.~Bigaran, B.~A.~Dobrescu and A.~Russo,
\href{doi:10.1103/PhysRevD.109.055033}{Phys. Rev. D \textbf{109}, no.5, 055033 (2024)}.
[arXiv:2312.09189 [hep-ph]].


\bibitem{ATLAS:2015qoy}
G.~Aad \textit{et al.} [ATLAS],
\href{doi:10.1007/JHEP09(2015)108}{JHEP \textbf{09}, 108 (2015)}.
[arXiv:1506.01291 [hep-ex]].

\bibitem{ATLAS:2023sbu}
G.~Aad \textit{et al.} [ATLAS],
\href{doi:10.1007/JHEP07(2023)118}{JHEP \textbf{07}, 118 (2023)}.
[arXiv:2303.05441 [hep-ex]].

\bibitem{CMS:2019hsm}
A.~M.~Sirunyan \textit{et al.} [CMS],
\href{doi:10.1103/PhysRevD.100.052003}{Phys.\ Rev.\ D \textbf{100}, 052003 (2019)}.
[arXiv:1905.10853 [hep-ex]].

\bibitem{CMS:2022nty}
A.~Tumasyan \textit{et al.} [CMS],
\href{doi:10.1103/PhysRevD.105.112007}{Phys.\ Rev.\ D \textbf{105}, 112007 (2022)}.
[arXiv:2202.08676 [hep-ex]].

\bibitem{CMS:2022cpe}
A.~Tumasyan \textit{et al.} [CMS],
\href{doi:10.1016/j.physletb.2023.137713}{Phys.\ Lett.\ B \textbf{846}, 137713 (2023)}.
[arXiv:2208.09700 [hep-ex]].

\bibitem{CMS:2024bni}
A.~Hayrapetyan \textit{et al.} [CMS],
\href{doi:10.1016/j.physrep.2024.09.012}{Phys.\ Rep.\ \textbf{1115}, 570-677 (2025)}.
[arXiv:2405.17605 [hep-ex]].
\bibitem{ATLAS:2024mrr}
G.~Aad \textit{et al.} [ATLAS],
\href{doi:10.1007/JHEP05(2025)075}{JHEP \textbf{05}, 075  (2025)}.

\bibitem{Bissmann:2020lge}
S.~Bi\ss{}mann, G.~Hiller, C.~Hormigos-Feliu and D.~F.~Litim,
\href{doi:10.1140/epjc/s10052-021-08886-3}{Eur.\ Phys.\ J.\ C \textbf{81}, 101 (2021)}.
[arXiv:2011.12964 [hep-ph]].

\bibitem{Bhattiprolu:2019vdu}
P.~N.~Bhattiprolu and S.~P.~Martin,
\href{doi:10.1103/PhysRevD.100.015033}{Phys.\ Rev.\ D \textbf{100}, 015033 (2019)}.
[arXiv:1905.00498 [hep-ph]].

\bibitem{ILC:2013jhg}
H.~Baer \textit{et al.} [ILC],
[arXiv:1306.6352 [hep-ph]].

\bibitem{ILCInternationalDevelopmentTeam:2022izu}
A.~Aryshev \textit{et al.} [ILC International Development Team],
[arXiv:2203.07622 [physics.acc-ph]].

\bibitem{CLICDetector:2013tfe}
H.~Abramowicz \textit{et al.} [CLIC Detector and Physics Study],
[arXiv:1307.5288 [hep-ex]].

\bibitem{Franceschini:2019zsg}
R.~Franceschini,
\href{doi:10.1142/S0217751X20410158}{Int.\ J.\ Mod.\ Phys.\ A \textbf{35}, 2041015 (2020)}.
[arXiv:1902.10125 [hep-ph]].



\bibitem{Yang:2021dtc}
B.~Yang, J.~Li, M.~Wang and L.~Shang,
\href{doi:10.1103/PhysRevD.104.055019}{Phys.\ Rev.\ D \textbf{104}, 055019 (2021)}.


\bibitem{Shang:2021mgn}
L.~Shang, M.~Wang, Z.~Heng and B.~Yang,
\href{doi:10.1140/epjc/s10052-021-09152-2}{Eur.\ Phys.\ J.\ C \textbf{81}, 415 (2021)}.

\bibitem{Bhattacharya:2021ltd}
S.~Bhattacharya, S.~Jahedi and J.~Wudka,
\href{doi:10.1007/JHEP05(2022)009}{JHEP \textbf{05}, 009 (2022)}.
[arXiv:2106.02846 [hep-ph]].
\bibitem{Shang:2023rfv}
L.~Shang, J.~Li, X.~Jia and B.~Yang,
\href{doi:10.1016/j.nuclphysb.2022.116071}{Nucl.\ Phys.\ B \textbf{987}, 116071 (2023)}.


\bibitem{Bhattiprolu:2023yxa}
P.~N.~Bhattiprolu, S.~P.~Martin and A.~Pierce,
\href{doi:10.1103/PhysRevD.109.035009}{Phys.\ Rev.\ D \textbf{109}, 035009 (2024)}.
[arXiv:2308.08386 [hep-ph]].

\bibitem{Yue:2024sds}
C.~X.~Yue, Y.~Q.~Wang, H.~Wang, Y.~H.~Wang and S.~Li,
\href{doi:10.1016/j.nuclphysb.2024.116482}{Nucl.\ Phys.\ B \textbf{1000}, 116482 (2024)}.
[arXiv:2402.02072 [hep-ph]].

\bibitem{Yue:2024ftz}
C.~X.~Yue, Y.~Q.~Wang, X.~C.~Sun and X.~Y.~Li,
\href{doi:10.1088/1361-6471/ad9ec8}{J. Phys. G \textbf{52}, no.2, 025003 (2025)}.
[arXiv:2412.07125 [hep-ph]].

\bibitem{Shen:2025mxe}
J.~F.~Shen, Y.~J.~Zhang and L.~Han,
\href{doi:10.1088/1674-1137/adf1ef}{Chin. Phys. C \textbf{49}, no.11, 113108  (2025)}.

\bibitem{Li:2025epjc}
R.~P.~Li, J.~W.~Lian and Y.~B. Liu,
 \href{doi:10.1140/epjc/s10052-025-15178-7}{Eur.\ Phys.\ J.\ C \textbf{85}, 1429 (2025)}.
\bibitem{ATLAS:2019gqq}
M.~Aaboud \textit{et al.} [ATLAS],
\href{doi:10.1103/PhysRevD.99.092007}{Phys.\ Rev.\ D \textbf{99}, 092007 (2019)}.
[arXiv:1902.01636 [hep-ex]].

\bibitem{mg5}
J.~Alwall, R.~Frederix, S.~Frixione, V.~Hirschi, F.~Maltoni, O.~Mattelaer, H.-S.~Shao, T.~Stelzer, P.~Torrielli and M.~Zaro,
\href{doi:10.1007/JHEP07(2014)079}{JHEP \textbf{07}, 079 (2014)}.
[arXiv:1405.0301 [hep-ph]].

\bibitem{pythia8}
T.~Sj\"ostrand, S.~Ask, J.~R.~Christiansen \textit{et al.},
\href{doi:10.1016/j.cpc.2015.01.024}{Comput.\ Phys.\ Commun.\ \textbf{191}, 159 (2015)}.
[arXiv:1410.3012 [hep-ph]].

\bibitem{deFavereau:2013fsa}
J.~de Favereau \textit{et al.} [DELPHES 3],
\href{doi:10.1007/JHEP02(2014)057}{JHEP \textbf{02}, 057 (2014)}.
[arXiv:1307.6346 [hep-ex]].
\bibitem{Leogrande:2019qbe}
E.~Leogrande, P.~Roloff, U.~Schnoor, and M.~Weber,
A DELPHES card for the CLIC detector,
\href{https://arxiv.org/abs/1909.12728}{arXiv:1909.12728}.


\bibitem{ma5}
E.~Conte, B.~Fuks and G.~Serret,
\href{doi:10.1016/j.cpc.2012.09.009}{Comput.\ Phys.\ Commun.\ \textbf{184}, 222 (2013)}.
[arXiv:1206.1599 [hep-ph]].

\bibitem{Conte:2014zja}
E.~Conte, B.~Dumont, B.~Fuks and C.~Wymant,
\href{doi:10.1140/epjc/s10052-014-3103-0}{Eur.\ Phys.\ J.\ C \textbf{74}, 3103 (2014)}.
[arXiv:1405.3982 [hep-ph]].

\bibitem{Cowan:2010js}
G.~Cowan, K.~Cranmer, E.~Gross and O.~Vitells,
\href{doi:10.1140/epjc/s10052-011-1554-0}{Eur.\ Phys.\ J.\ C \textbf{71}, 1554 (2011)}.
[Erratum: Eur.\ Phys.\ J.\ C \textbf{73}, 2501 (2013)].
[arXiv:1007.1727 [physics.data-an]].

\end{thebibliography}
\end{document}